\documentclass[twocolumn]{bmcart}

\RequirePackage{natbib}
\RequirePackage{hyperref}
\usepackage[utf8]{inputenc} 
\usepackage{amsmath, amsthm, amssymb}
\usepackage{graphicx}
\usepackage{booktabs}
\usepackage{multirow}
\usepackage{graphicx}

\begin{document}

  \begin{frontmatter}

  \begin{fmbox}
  \dochead{Research}
  \title{Multi-Objective De Novo Drug Design with Conditional Graph Generative Model}

  \author[]{\inits{Y}\fnm{Yibo} \snm{Li}}
  \author[]{\inits{L}\fnm{Liangren} \snm{Zhang*}}
  \author[]{\inits{Z}\fnm{Zhenming} \snm{Liu*}}

  *correspondence: \url{liangren@bjmu.edu.cn}; \url{zmliu@bjmu.edu.cn}

  State Key Laboratory of Natural and Biomimetic Drugs, School of Pharmaceutical Sciences, Peking University, Xueyuan Road 38, Haidian District, 100191, Beijing, China

  \begin{abstractbox}

    \begin{abstract}
      Recently, deep generative models have revealed itself as a promising way of performing \textit{de novo} molecule design. However, previous research has focused mainly on generating SMILES strings instead of molecular graphs. Although current graph generative models are available, they are often too general and computationally expensive, which restricts their application to molecules with small sizes. In this work, a new \textit{de novo} molecular design framework is proposed based on a type sequential graph generators that do not use atom level recurrent units. Compared with previous graph generative models, the proposed method is much more tuned for molecule generation and have been scaled up to cover significantly larger molecules in the ChEMBL database. It is shown that the graph-based model outperforms SMILES based models in a variety of metrics, especially in the rate of valid outputs. For the application of drug design tasks, conditional graph generative model is employed. This method offers higher flexibility compared to previous fine-tuning based approach and is suitable for generation based on multiple objectives. This approach is applied to solve several drug design problems, including the generation of compounds containing a given scaffold, generation of compounds with specific drug-likeness and synthetic accessibility requirements, as well as generating dual inhibitors against JNK3 and GSK3$\beta$. Results show high enrichment rates for outputs satisfying the given requirements.
    \end{abstract}


  \end{abstractbox}
  \end{fmbox}
  \end{frontmatter}

  \section*{Introduction}

    The ultimate goal of drug design is the discovery of new chemical entities with desirable pharmacological properties. Achieving this goal requires medicinal chemists to perform searching and optimization inside the space of new molecules. This task is proved to be extremely difficult, mainly due to the size and complexity of the search space. It is estimated that there are around $10^{60}\sim 10^{100}$ synthetically available molecules\cite{review_denovo_1}. Meanwhile, the space of chemical compounds exhibits a discontinues structure, making searching difficult to perform\cite{svae_1}.

    {\it De novo} molecular design aims at assisting this processes with computer-based methods. Early works have developed various algorithms to produce new molecular structures, such as atom based elongation or fragment based combination\cite{ludi, fragment}. Those algorithms are often coupled with global optimization techniques such as ant colony optimization\cite{aco_1, aco_2}, genetic algorithms\cite{ga_1,ga_2} or particle swam optimization\cite{pso} for the generation of molecules with desired properties. 

    Recent developments in deep learning\cite{book_dl} have shed new light on the area of {\it de novo} molecule  generation. Works have shown that deep generative models are very effective at modeling the SMILES representation of molecules using recurrent neural networks (RNN), an architecture that  has been extensively applied to tasks related sequential data\cite{review_rnn}. Segler et al\cite{slm_3} applied SMILES language model (LM) to the task of  generating focused molecule libraries by fine-tuning the trained network with a smaller set  of molecules with desirable properties. Olivecrona et al\cite{slm_1} used  a GRU\cite{gru} based LM trained on the ChEMBL\cite{chembl} dataset to generate SMILES string. The mode is then fine-tuned using reinforcement learning for the generation of molecules with  specific requirements. Popova et al\cite{slm_2} propose to integrate the generative and predictive  network together in the generation phase. Segler et al\cite{slm_3} applied SMILES LM to the task of  generating focused molecule libraries by fine-tuning the trained network with a smaller set  of molecules with desirable properties. Beside language model, Gómez-Bombarelli et al\cite{slm_1} used variational autoencoder (VAE)\cite{vae} to generate drug-like compounds from ZINC database\cite{zinc}. This  work aims at obtaining a bi-directional mapping between molecule space and a continuous latent space so that operations on molecules can be achieved by manipulating the latent  representation. Blaschke et al\cite{svae_2} compared different architectures for VAE and applied it  to the task of designing active compounds against DRD2.

    The works described above demonstrated the effectiveness of SMILES based model regarding  molecule generation. However, producing valid SMILES strings requires the model to learn rules  that are irrelevant to molecular structures, such as the SMILES grammar and atom ordering, which adds unnecessary burden to the training process, making the SMILES string a less  preferable representation compared to molecular graphs. Research in deep learning has recently  enabled the direct generation of molecular graphs. Johnson et al\cite{glm_1} proposed a sequential  generation approach for graphs. Though their implementation is mainly for reasoning tasks,  this framework provided is potentially applicable to molecule generation. Compared with  this approach, a more recent method\cite{gvae_1} was proposed for generating the entire graph all at  once. This model has been successfully applied to the generation of small molecular  graphs. The implementation that is most similar to ours is by the recent work by Li et al\cite{glm_2} using a sequential decoding scheme similar to that by Johnson et al. Decoding invariance is  introduced by sampling different atom ordering from a predefined distribution. This method  has been applied to the generation of molecules with less than 20 heavy atoms from ChEMBL  dataset. Though inspiring, the methods discussed above have a few common problems. First of  all, the generators proposed are relatively general. This design allows those techniques to  be applied to various scenarios but requires further optimization for application in  molecule generation. Secondly, many of those models suffer from scalability issue, which  restricts the application to molecules with small sizes.

    \begin{figure}[t!]
        \includegraphics{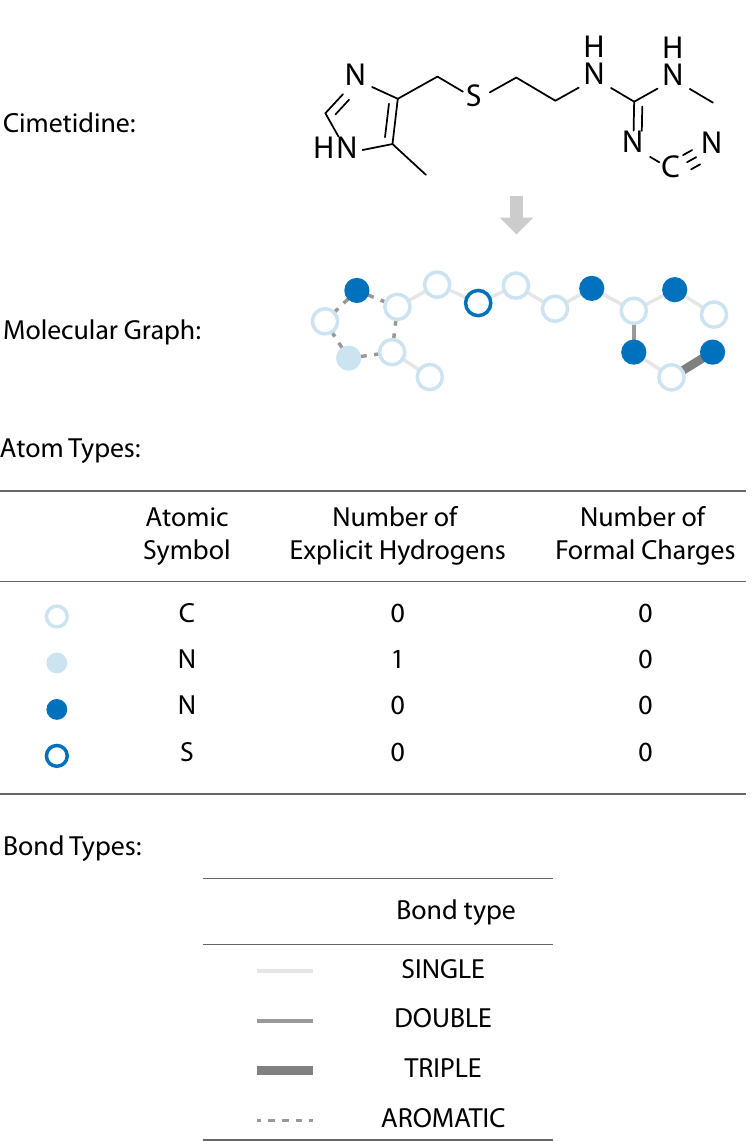}
        \caption{
            \csentence{Cimetidine and its graph based representation}
            In the graph based generative models, molecules are represented as graphs $G=(V, E)$, where atoms are bonds are viewed as nodes and edges respectively (see {\bf a} and {\bf b}). Atom types are specified by three parameters: the atomic symbol (or equally the atomic number), the number of explicit hydrogens attached, and the number of formal charges (see {\bf c}). For bond types, only single, double, triple and aromatic bonds are considered in this work (see {\bf d}).
        }
        \label{fig:molecule_representation}
    \end{figure}

    In this work, we propose a graph-based generator that is more suited for molecules.  The model is scaled to cover compounds containing up to 50 heavy atoms in the ChEMBL dataset.  Results show the graph-based model proposed is able to outperform SMILES based methods in a variety metrics, including the rate of valid outputs, KL and JS divergence of molecular properties, as well as NLL loss. A conditional version of the model is employed to solve  various drug design related tasks with multiple objectives, and promising performance has  been demonstrated according to the results.

    \begin{figure*}
        \includegraphics{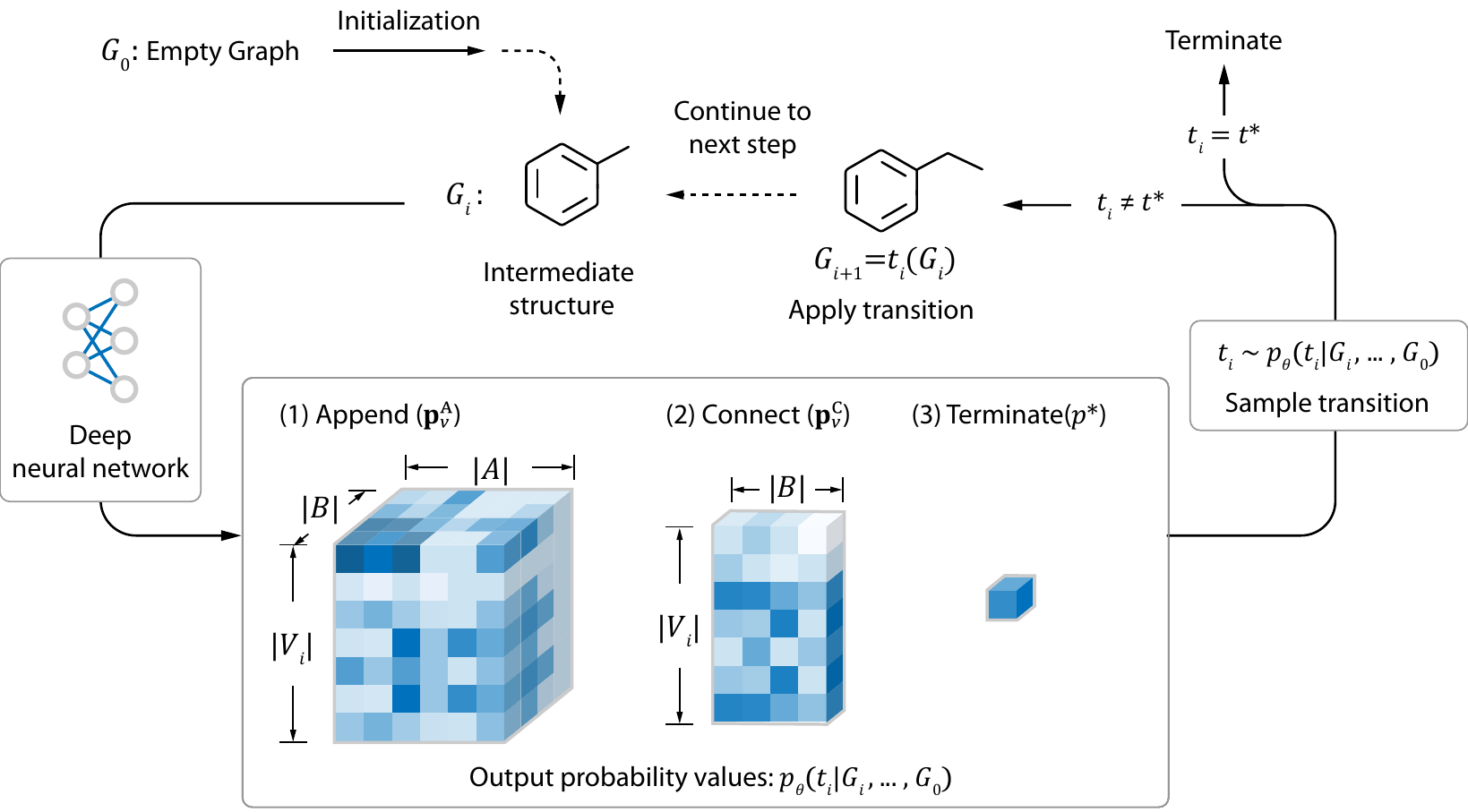}
        \caption{
            \csentence{A schematic representation of molecule generation process}
            Starting with the empty graph $G_0$, initialization is performed to add the first atom. At each step, a graph transition (append, connect or terminate) is sampled and performed on the intermediate molecule structure. The probability for sampling each transition is given by $p_{\boldsymbol \theta}(t|G_i,..., G_0)$, which is parametrized using deep neural network. Finally, termination operation is performed to end the generation. 
        }
        \label{fig:decoding_process}
    \end{figure*}

  \section*{Methods}
    \subsection*{Molecular Graph}
      Molecular graph is a way of representating the structural information of molecules using graph objects ($G=(V, E)$), where atoms and bonds as viewed as graph nodes ($v \in V$) and edges ($e \in E$). Each node $v$ in $V$ is labeled with its corresponding atom type. In this work, the atom type is specified using three variables: the atomic symbol (or equally the atomic number), the number of explicit hydrogens attached, and the number of formal charges. For example, the nitrogen atom in pyrrole can be represented as the triple (``N'', 1, 0). Similarly, the edges in $E$  are labeled with bond types. Only four types of bonds are considered in this work: single, double, triple and aromatic.

      The set of all atom types and all bond types are denoted as $A$ and $B$ respectively. $A$ is extracted from molecules in the ChEMBL dataset (see Supplementary Text 1), and contains 33 elements in total. A visualized demonstration of molecular graph is given in Figure \ref{fig:molecule_representation}.

      \begin{figure*}
        \includegraphics{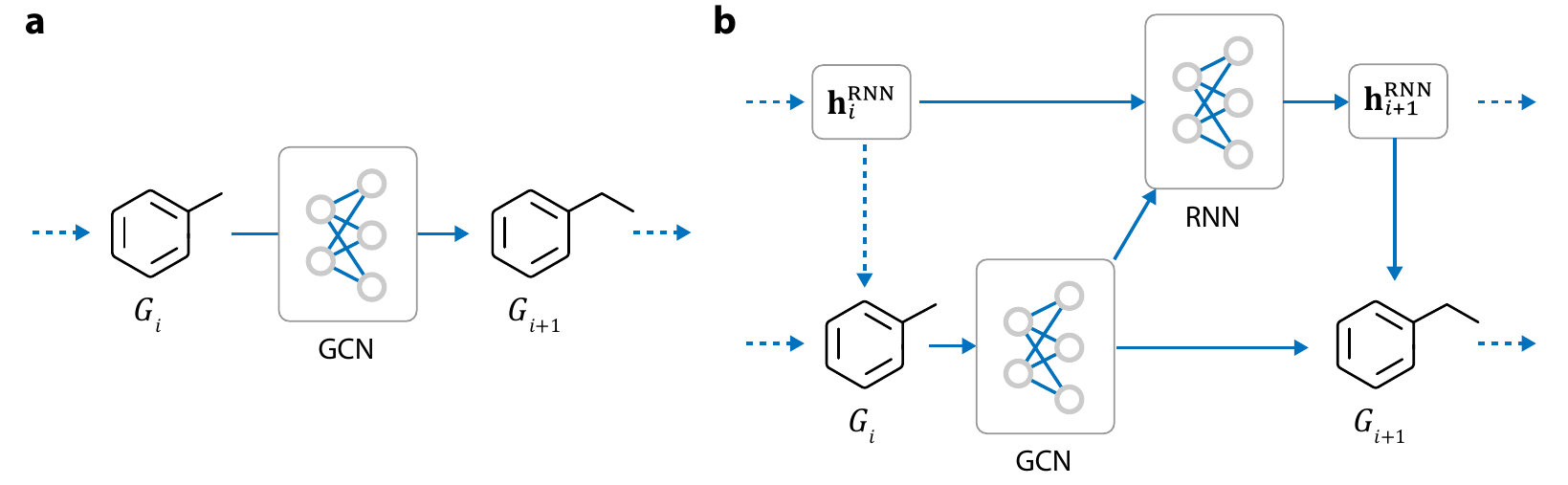}
        \caption{
            \csentence{The two type of graph generative architectures explored in this work} {\bf a}. MolMP: This architecture models graph generation as a Markov process, where the transition of $G_i$ only depends on the current state of the graph, not on the history. {\bf b}. MolRNN: This architecture adds a single molecule level recurrent unit to MolMP.
        }
        \label{fig:recurrency}
      \end{figure*}

    \subsection*{Graph Generative Model}
      We now consider the deep generative models that can directly output molecular graphs. In this work, we mainly focus on sequential graph generators, which builds graph by iteratively refining its intermediate structure. The process starts from the empty graph $G_0=(\emptyset,\emptyset)$. At step $i$, a graph transition $t_i$ is selected from the set of all available transition actions $T(G_i)$ based on the generation history $(G_0 , ..., G_i)$. The selection is done by sampling $t_i$ from a probability distribution $t_i \sim p_{\boldsymbol \theta}(t_i | G_i, ..., G_0)$ parametrized by neural network. Then, $t_i$  is performed on $G_i$ to get the graph structure for the next step $G_{i+1}=t_i (G_i )$.  At the final step $n$, termination operation $t^*$ is performed and the model outputs $G=G_n$ as the final product.

      The entire process is illustrated in Figure \ref{fig:decoding_process}. We call the mapping $T$, which determines all available graph transitions at each step, a \textit{decoding scheme}. The sequence  $r=((G_0,t_0),(G_1,t_1),...,(G_n,t_n))$ is called a \textit{decoding route} of $G$, and the distribution $p_{\boldsymbol \theta} (t_i | G_i, ..., G_0)$  is called a \textit{decoding policy}.

      Previous graph generative models are usually too general and less optimized for the generation of molecular graphs. Here we offer the following optimizations:

      \begin{enumerate}
          \item A much simpler decoding scheme $T$ is used to decrease the number of steps required for generation.

          \item No atom level recurrent unit is used in the decoding policy. Instead, we explored two other options: (1) parametrizing the decoding policy as a Markov process and (2) using only molecule level recurrent unit. Those modifications helps to increase the scalability of the model.  

          \item During the calculation of log-likelihood loss, we sample $r$ from a parametrized distribution $q_\alpha (r | G)$. The parameter $\alpha$ controls the degree of randomness of $q_\alpha$, offering higher flexibility for the model.

      \end{enumerate}

      The following three sections are devoted to the detailed discussions of the optimizations above.

    \subsection*{Decoding Scheme}
      The transitions in $T(G_i)$ given the intermediate state $G_i$ is restricted to the following four types:

      \begin{enumerate}
          \item {\bf Initialization}: At the beginning of the generation, the only allowed transition is to add the first atom to the empty graph $G_0$.

          \item {\bf Append}: This action adds a new atom to $G_i$ and connect it to an existing atom with a new bond. 

          \item {\bf Connect}: This action connects two existing atoms $v_1,v_2 \in V_i$ with a new bond. For simplicity, we only allow connections to start from the latest appended atom $v^*$, which means that $v_1=v^*$.

          \item {\bf Termination}: End the generation process. This action is denoted as $t^*$.

      \end{enumerate}

      The entire process is shown in Figure \ref{fig:decoding_process}, and a more detailed illustration is provided in Figure S1 and S2(Additional file 2). In theory, $T(G)$ should not contain actions that violate the validity constraints of molecules. However, in order to test the ability for the model to learn those constraints, we do not explicity exclude those actions from $T(G)$ during training. 

      Note that compared with the implementation in \cite{glm_2}, the action of adding new atom and the action of connecting it to the molecule is merged into a single ``append'' step. This helps to reduce the number of steps during generation. It is easy to show that the number of steps required for generating graph $G=(V, E)$ equals exactly to $|E|+2$, which is generally much smaller than the length of the corresponding SMILES string (as shown in Figure S3(Additional file 2)).

      \subsection*{Decoding Policy}
      During generation, the decoding policy $p_{\boldsymbol \theta}$ need to specify the probability value for each graph transition in $T(G_i)$. More specifically, $p_{\boldsymbol \theta}$ need to output the following probability values:

      \begin{enumerate}
          \item ${\bf p}_v^\textrm{A}$ for each $v \in V_i$: A matrix with size $|A|\times|B|$, whose element $({\bf p}_v)_{ab}$ represents the probability of appending a new atom of type $a \in A$ to atom $v$ with a new bond of type $b \in B$.

          \item ${\bf p}_v^\textrm{C}$ for each $v \in V_i$: A vector with size $|B|$, whose element $({\bf p}_v^\textrm{C})_{b}$ represents the probability of connecting the latest added atom $v^*$ with $v$ using a new bond of type $b\in B$.

          \item $p^*$: A scalar value indicating the probability of terminating the generation.
      \end{enumerate}

      A visualized depiction of ${\bf p}_v^\textrm{A}$, ${\bf p}_v^\textrm{C}$ and $p^*$ is shown in Figure \ref{fig:decoding_process}. The decoding policy $p_{\boldsymbol \theta}$ is parameterized using neural network. At each step, the network accepts the the decoding history $(G_0, ..., G_i)$ as input and calculates the probability values (${\bf p}_v^\textrm{A}$, ${\bf p}_v^\textrm{C}$, $p^*$) as output. In this work, we explored two novel graph generation architectures, namely MolMP and MolRNN. Unlike the methods proposed in \cite{glm_1, glm_2}, the two architectures do not involve atom level recurrency, which helps to increase the scalability of the model.

      \subsubsection*{MolMP}

          The first architecture models graph generation as a Markov process, where the transition of $G_i$ only depends on the current state of the graph, not on the history (Figure \ref{fig:recurrency}{\bf a}). This means that $p_{\boldsymbol \theta}(t|G_i, ..., G_0)=p_{\boldsymbol \theta}(t|G_i)$. We refer to this method as MolMP. Since this type of architecture does not include any recurrent units, it will be less expensive compared with RNN based models. Moreover, the computation at different steps can be easily parallelized during training. The detailed architecture of MolMP is given as follows:

          \begin{enumerate}
              \item An initial atom embedding ${\bf h}^0_v$ is first generated for each atom $v$: \\
                  \begin{equation}
                      {\bf h}^0_v = \textrm{Embedding}_{\boldsymbol \theta}(v)
                  \end{equation}
                  ${\bf h}^0_v$ is determined based on the following information: (1) the atom type of $v$ and (2) whether $v$ is the latest appended atom. The dimension of ${\bf h}^0_v$ is set to 16.

              \item ${\bf h}^0_v$ is passed to a sequence of $L$ graph convolutional layers: \\
                  \begin{equation}
                      {\bf h}^l_v = \textrm{GraphConv}^l_{\boldsymbol \theta}({\bf h}^{l-1}_v, G_i)
                  \end{equation}
                  Where $l=1, ..., L$. The outputs from all graph convolutional layers are then concatenated together, followed by batch normalization and ReLU:\\
                  \begin{equation}
                      {\bf h}_v^\textrm{skip} = relu(bn(\textrm{Concat}({\bf h}^1_v, ..., {\bf h}^L_v)))
                  \end{equation}
                  Except the first layer, each convolutional layer $\textrm{GraphConv}^l_{\boldsymbol \theta}$ adopts a ``BN-ReLU-Conv'' structure as suggested in \cite{resnet}. The detailed architecture of graph convolution is described in \hyperref[graph-conv]{``Graph Convolution''}. We use six convolution layers in this work ($L=6$), each with 32, 64, 128, 128, 256, 256 output units. 

              \item ${\bf h}_v^\textrm{skip}$ is passed to the fully connected network $\textrm{MLP}^\textrm{FC}_{\boldsymbol \theta}$ to obtain the final atom level representation ${\bf h}_v$.\\
                  \begin{equation}
                      {\bf h}_v = \textrm{MLP}^\textrm{FC}_{\boldsymbol \theta}({\bf h}_v^\textrm{skip})
                  \end{equation}
                  $\textrm{MLP}^\textrm{FC}_{\boldsymbol \theta}$ consists of two linear layers, with 256 and 512 output units each. Batch normalization and ReLU are applied after each layer.

              \item Average pooling is applied to obtain the molecule level representation ${\bf h}_{G_i}$:\\
                  \begin{equation}
                      {\bf h}_{G_i}=AvgPool([{\bf h}_v]_{v\in V_i})
                  \end{equation}

              \item The activation value for each transition in $T(G_i)$ is obtained using ${\bf h}_v$ and ${\bf h}_{G_i}$. \\
                  \begin{align}
                      &{\bf s}_v=\textrm{MLP}_{\boldsymbol \theta}({\bf h}_v, {\bf h}_{G_i}) \label{calc_activation_1}\\
                      &s^*=\textrm{MLP}^*_{\boldsymbol \theta}({\bf h}_{G_i}) \label{calc_activation_2}
                  \end{align}
                  For each atom $v \in V_i$, ${\bf s}_v$ is a matrix of size $|A|\times|B|+|B|$, which is subsequently split into ${\bf s}_v^\textrm{A}$ and ${\bf s}_v^\textrm{C}$ with size $|A|\times|B|$ and $|B|$respectively. $s^*$ is a scalar containing the activation value for termination action $t^*$. $\textrm{MLP}_{\boldsymbol \theta}$is a two layer fully connected network with hidden size 128. $\textrm{MLP}^*$ is a one layer fully connected network. Both $\textrm{MLP}_{\boldsymbol \theta}$ and $\textrm{MLP}^*$ uses exponential activiaton in the output layer.

              \item The activation values are normalized to give the probability values:\\
                  \begin{align}
                      &{\bf p}^\textrm{A}_v = {\boldsymbol s}_v^\textrm{A}/S\\
                      &{\bf p}^\textrm{C}_v = {\boldsymbol s}_v^\textrm{C}/S\\
                      &p^* =s^*/S
                  \end{align}
                  where $S=\sum_{vab}({\bf s}^\textrm{A}_v)_{ab}+\sum_{vb}({\bf s}^\textrm{C}_v)_b + s^*$
          \end{enumerate}

\begin{figure}[t!]
            \includegraphics{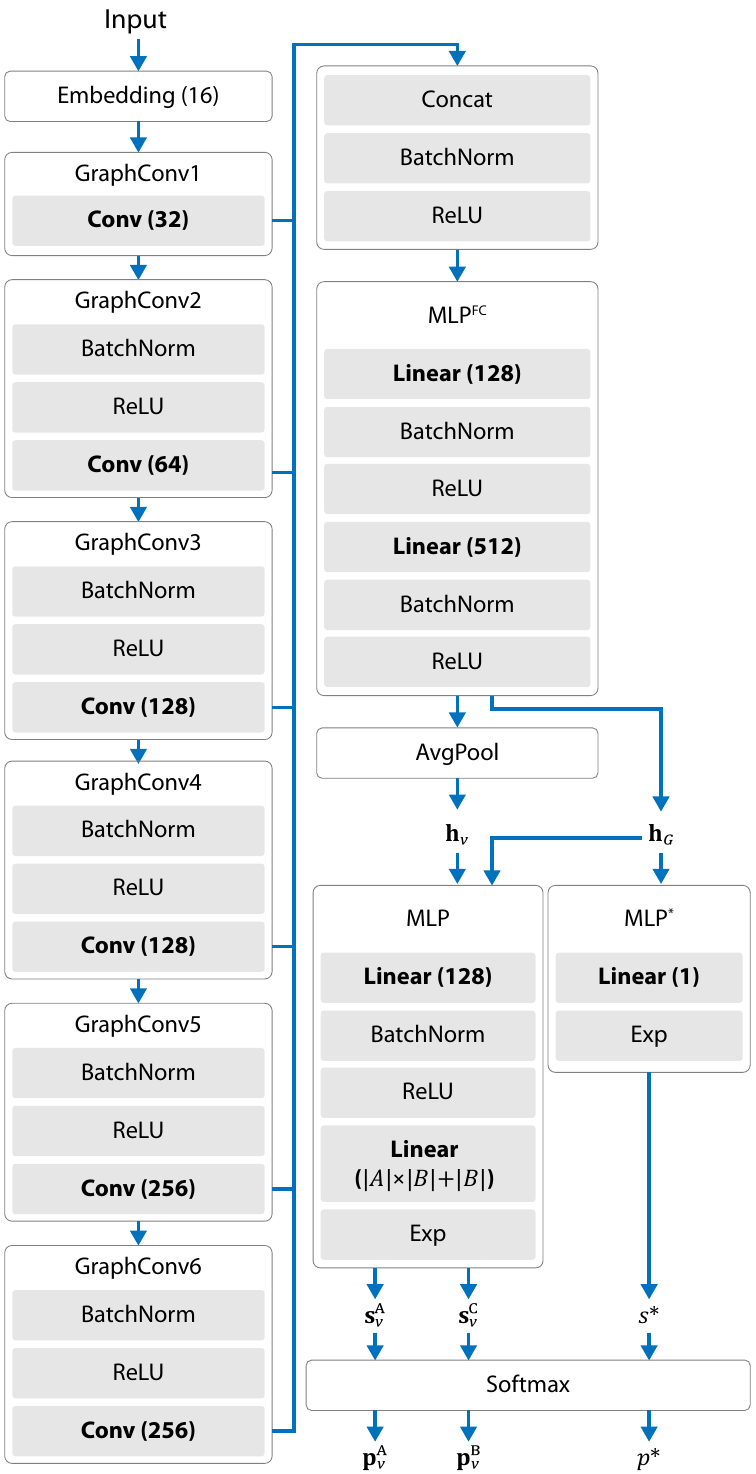}
            \caption{
                \csentence{Network architecture for MolMP} This figure shows the detailed model architecture for MolMP. MolRNN adopts a structure highly similar to that of MolMP, except the inclusion of the molecule level recurrent unit.
            }
            \label{fig:network_architecture}
          \end{figure}

          The architecture of the entire network is shown in Figure \ref{fig:network_architecture}.
    
          \begin{figure*}
            \includegraphics{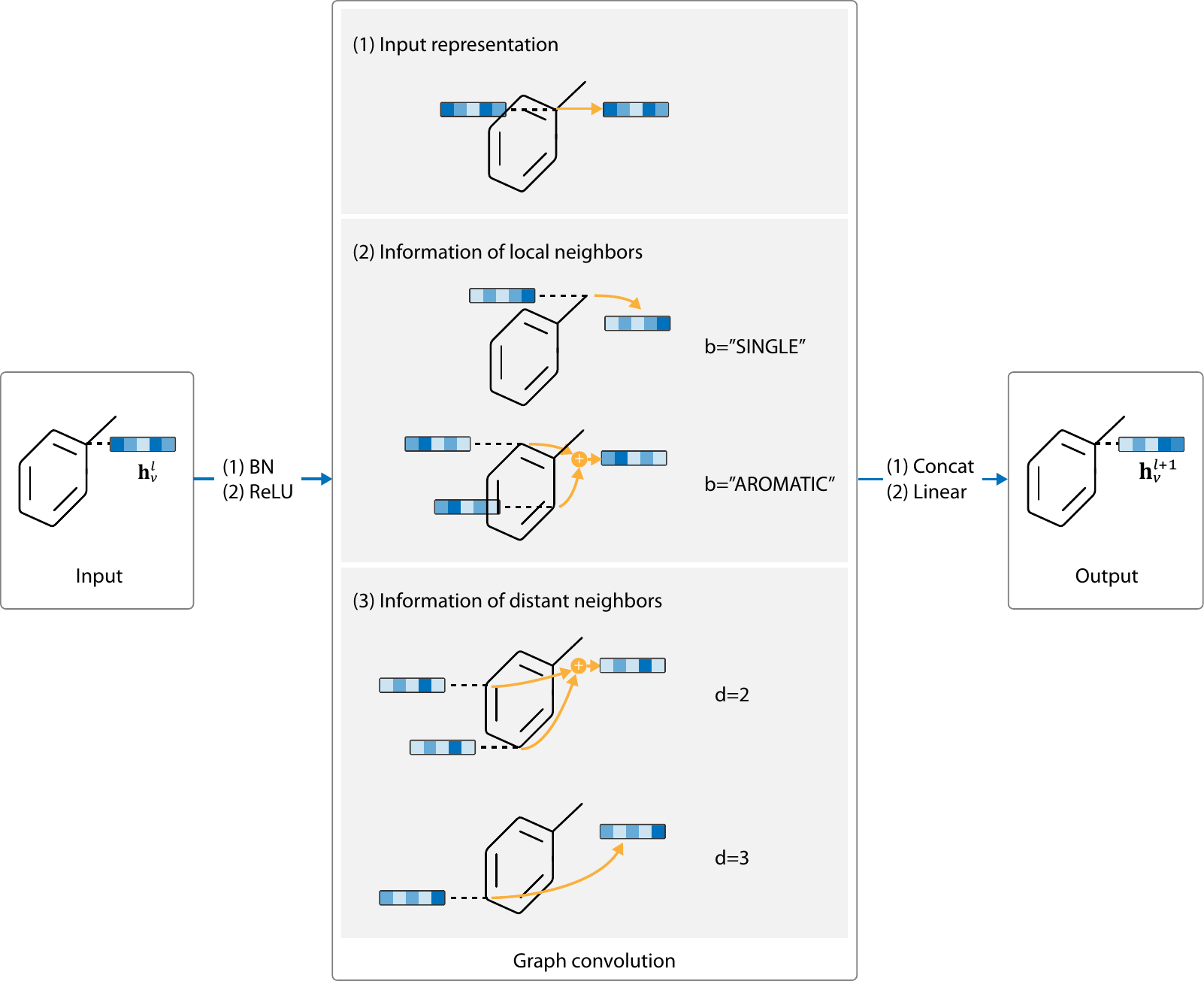}
            \caption{
                \csentence{Architecture of graph convolutional layer}
                At each layer, the output representation for atom $i$ is given by: (1) the input representation of $i$ from previous layers, (2) information of local neighbors and (3) information of distant neighbors.
            }
            \label{fig:graph_conv}
          \end{figure*}

      \subsubsection*{MolRNN}

          The second architecture adds a single molecule level recurrent unit to MolMP, as shown in Figure \ref{fig:recurrency}. We refer to this method as MolRNN. The model architecture is specified as follows:

          \begin{enumerate}

              \item First of all, the model generates the atom level (${\bf h}_v, v \in V_i$) and molecule level (${\bf h}_{G_i}$) representation for the graph state $G_i$. This part of the network uses the same architecture as that in MolMP.
          
              \item Given ${\bf h}_v$ and ${\bf h}_{G_i}$, the hidden state of the molecule level recurrent unit (${\bf h}_i^{RNN}$) is updated as:\\
              \begin{equation}
                  {\bf h}_{i+1}^\textrm{RNN}=\textrm{RNN}_{\boldsymbol \theta}({\bf h}_i^\textrm{RNN}, {\bf h}_{v*}, {\bf h}_{G_i})
              \end{equation}
              Where ${\bf h}_{v*}$ is the representation of the latest appended atom $v^*$. The recurrent network $\textrm{RNN}_{\boldsymbol \theta}$ is employed using three GRU layers with a hidden size of 512.
          
              \item The probability values ${\bf p}_v^\textrm{A}$, ${\bf p}_v^\textrm{C}$, $p^*$ are calculated in the same manner as MolMP by replacing ${\bf h}_{G_i}$ in eq. \ref{calc_activation_1} and eq. \ref{calc_activation_2} with ${\bf h}_{i+1}^\textrm{RNN}$.
          \end{enumerate}

          The overall architecture of MolRNN is highly similar to that of MolMP. However, it is found that the molecule level recurrent unit in MolRNN provides significant improvements to the model performance (see \hyperref[model-performance]{``Model Performance and Sample Quality''}), while inducing little extra computational cost compared with MolMP.

      \subsubsection*{Graph Convolution}

        \label{graph-conv}

        In this work, we rely on graph convolutional network (GCN)\cite{graph_conv_1} to extract information from graph states $G_i$. Each graph convolutional layer adopts the ``BN-ReLU-Conv'' structure as described before. In terms of the convolution part, the architecture is structured as follows:
        
        \begin{equation}
            \begin{split}
                {\bf h}_v^l = & {\bf W}^l{\bf h}_v^{l-1} + \\
                & \sum_{b\in B}{\Theta_b^l \sum_{u\in N_b^{bond}(v)}{{\bf h}_u^{l-1}}} + \\
                & \sum_{1 < d \le D}{\Phi_d^l \sum_{u\in N_d^{path}(v)}{{\bf h}_u^{l-1}}}
            \end{split}
            \label{unconditional-convolution}
        \end{equation}

        Where ${\bf h}_v^l$ is output representation of atom $v$ at layer $l$, and ${\bf h}_v^{l-1}$ is the input representation. $N_b^{bond} (v)$ is the set of all atoms directly connected to atom $v$ with bond of type $b$, and $N_d^{path} (v)$ is the set of all atoms whose distance to atom $v$ equals to $d$. $D$ represents the receptive field size, which is set to 3 in this work. $W^l$, $\Theta_b^l$ and $\Phi_d^l$ are weight parameters of layer $l$.

        Briefly speaking, at each layer $l$, the output representation of atom $v$ (${\bf h}_v^l$) is calculated according to the following information:

        \begin{enumerate}
            \item The input representation of $v$ (${\bf h}_v^{l-1}$),
            \item Information of local neighbors, which is given by $\sum_{b\in B}{\Theta_b^l \sum_{u\in N_b^{bond}(v)}{{\bf h}_u^{l-1}}}$. Note that this part of information is conditioned on the bond type $b$ between $v$ and its neighborhood atom $u$.
            \item Information of remote neighbors, given by \\
            $\sum_{1 < d \le D}{\Phi_d^l \sum_{u\in N_d^{path}(v)}{{\bf h}_u^{l-1}}}$. This part of information is conditioned on the distance $d$ between $v$ and its remote neighbor $u$.
        \end{enumerate}

        The architecture is illustrated in Figure \ref{fig:graph_conv}. Our implementation of graph convolution is similar to the edge conditioned convolution by Simonovsky el al\cite{graph_conv_2}, except that we also include the information of remote neighbors of $v$ in order to reach larger receptive field with fewer layers.

    \subsection*{Likelihood Function}
        To train the generative model, we need to maximize the log-likelihood $p_{\boldsymbol \theta} (G)$ for the training samples. However, for the step-wise generative models discussed above, the likelihood is only tractable for a given decoding route $r=((G_0,t_0),(G_1,t_1),...,(G_n,t_n))$:

        \begin{equation}
            \log{p_{\boldsymbol \theta}(G, r)} = \sum_{i=0}^{n}{\log{p_{\boldsymbol \theta}(t_i|G_i, ..., G_0)}}
        \end{equation}

        While the marginal likelihood can be computed as:

        \begin{equation}
            \log{p_{\boldsymbol \theta}(G)=\log{\sum_{r \in R(G)}{p_{\boldsymbol \theta}(G, r)}}}
        \end{equation}

        Where $R(G)$ is the set of all possible decoding route for $G$. The marginal likelihood function is intractable for most molecules encountered in drug design. One way to resolve this problem is to use importance sampling as proposed in \cite{glm_2}:

        \begin{equation}
            \log{p_{\boldsymbol \theta}(G)} = \log{\mathbb{E}_{r\sim q(r|G)}[\frac{p_{\boldsymbol \theta}(G, r)}{q(r|G)}]}
        \end{equation}

        Where $q(r|G)$ is a predefined distribution on $R(G)$. Both the deterministic and the fully randomized $q(r|G)$  were explored in the previous work\cite{glm_2}. However, a more desirable solution would lie in somewhere between  deterministic decoding and fully randomized decoding. In this work, instead of sample from the distribution $q(r|G)$,  we sample r from distribution $q_\alpha (r|G)$ that is parameterized by $0 \le \alpha \le 1$. $q_\alpha (r|G)$ is  designed such that the decoding will largely follow depth first decoding with canonical ordering, but at each step,  there is a small possibility $1-\alpha$ that the model will make a random mistake. In this way, the parameter $\alpha$ measures can be used to control the randomness of the distribution $q_\alpha$. The algorithm is shown in  Supplementary Text 4(Additional file 1).

        \begin{equation}
            \begin{split}
                \log{p_{\boldsymbol \theta}(G)}
                & = \log{\mathbb{E}_{r\sim q_\alpha(r|G)}[\frac{p_{\boldsymbol \theta}(G, r)}{q_\alpha(r|G)}]} \\
                & \ge \log{\frac{1}{k}\sum_{i=1}^k{\frac{p_{\boldsymbol \theta}(G, r_i)}{q_\alpha(r_i|G)}}}
            \end{split}
        \end{equation}

        For $\alpha=1$, the distribution falls back to the deterministic decoding. The parameter $\alpha$ is treated as a hyperparameter which is optimized for model performance. We tried $\alpha \in \{1.0,0.8,0.6\}$ on both MolMP and MolRNN.
        \subsection*{Conditional Generative Model}

\begin{figure}[h!]
            \includegraphics{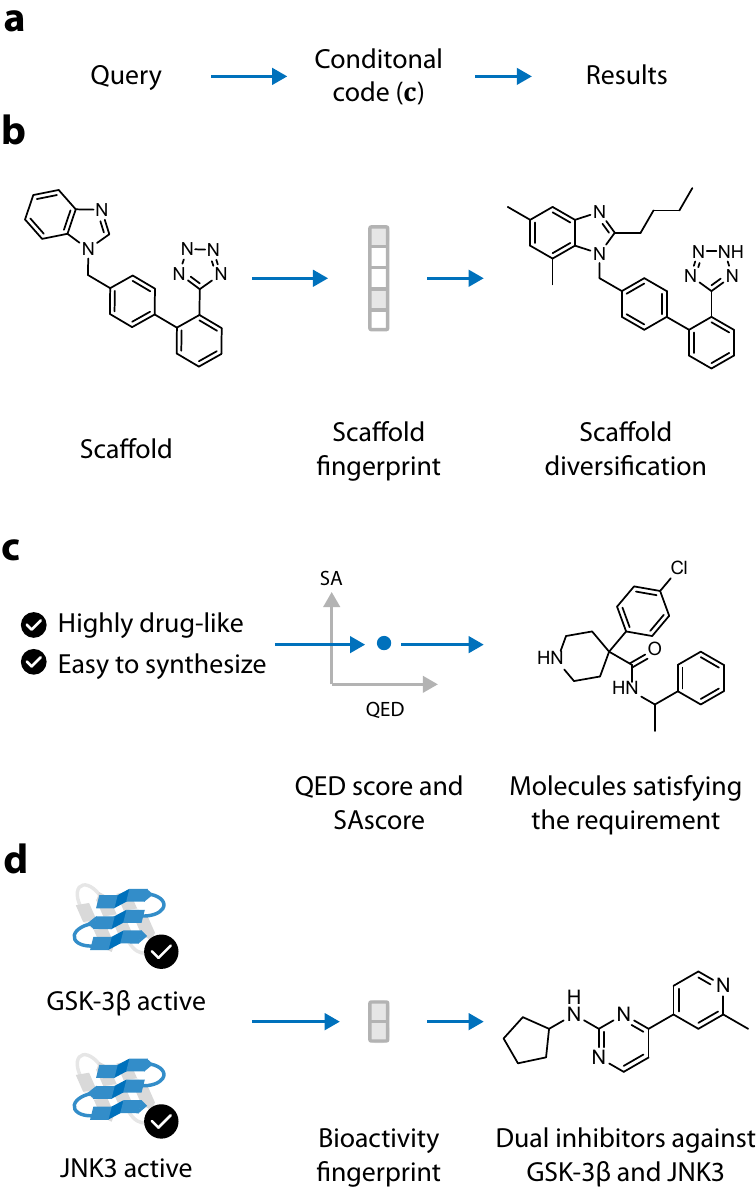}
            \caption{
                \csentence{Conditonal generative models} 
                {\bf a}. For the generation of molecules based on requriements, the requriement(query) is first converted to the numerical representation called conditoinal code $\bf c$, the generative model is then modified to be conditioned on $\bf c$.
                {\bf b}. Scaffold based molecule generation. 
                {\bf c}. Generation based on drug-likeness and synthetic accessibility.
                {\bf d}. Designing of dual inhibitors of JNK3 and GSK-3$\beta$ 
            }
            \label{fig:conditional_models}
        \end{figure}

        Most molecule design tasks require to produce compounds satisfying certain criteria, such as being synthetically  available or having a high affinity for a certain target. Currently, the most popular  solution is to fine-tune the existing model so that it can be suited for a specific task\cite{slm_1, slm_2, slm_3}. However, modeling multiple objectives is challenging for this type of models.  Herein, conditional generative model is propose for generation tasks with specific requirements. We first convert the given requirement to the numerial representation called conditional code ($\bf c$), and the generative model is then modified to be conditioned on $\bf c$. For graph generative model, this means that the decoding policy is now $p_{\boldsymbol \theta} (t_i|G_i, ..., G_0, {\bf c})$ (see Figure \ref{fig:conditional_models}). Compared with fine-tuning based methods, conditional model can be easily applied to multi-objective and multi-task settings.

        Both graph based and SMILES based conditional generators are implemented in this work. For graph based model, the graph convolution is modified to include $\bf c$ as input:
        \begin{equation}
            \begin{split}
                {\bf h}_v^l = & {\bf W}^l{\bf h}_v^{l-1} + \\
                & \sum_{b\in B}{\Theta_b^l \sum_{u\in N_b^{bond}(v)}{{\bf h}_u^{l-1}}} + \\
                & \sum_{1 < d \le D}{\Phi_d^l \sum_{u\in N_d^{path}(v)}{{\bf h}_u^{l-1}}} + \Psi^l{\bf c}
            \end{split}
        \end{equation}
        Simply state, $\bf c$ is included in the graph convoludion architecture by adding an additional term $\Psi^l{\bf c}$ to the unconditional implementation in eq. \ref{unconditional-convolution}. For SMILELS based model, the conditional code is included by concatenating it with the input at each step: ${\bf x}_i^\prime=\textrm{Concat}({\bf x}_i, {\bf c})$. Where ${\bf x}_i$ is the one-hot representation of the SMILES charactor input at step $i$. 
    
        Conditional models have already been used by the previous work\cite{gvae_1} for molecule generation, but was restricted to small molecules and have only used simple properties such as the number of heavy atoms as conditional codes. Here, the model is applied to tasks that are much more related to drug design, including scaffold-based generation, property-based generation and the design of dual inhibitor of JNK3 and GSK-3$\beta$ (see \ref{fig:conditional_models}).
    
    \subsection*{Scaffold-Based Generation}
        The concept of molecular scaffold has long been of significant importance in medicinal chemistry\cite{review_scaffold}. Though various definitions are available, the most widely accepted definition is given by Bemis and Murcko\cite{bemis-murcko}, who proposed derive the scaffold of a given molecule by removing all side chain atoms. Studies have found various scaffolds that have privileged characteristics in terms of the activity of certain target\cite{scaffold_1,scaffold_2,scaffold_3}. Once such privileged structure is found, a related task is to produce compound libraries containing such scaffolds for subsequent screening. 
        
        \begin{figure*}
            \includegraphics[scale=0.9]{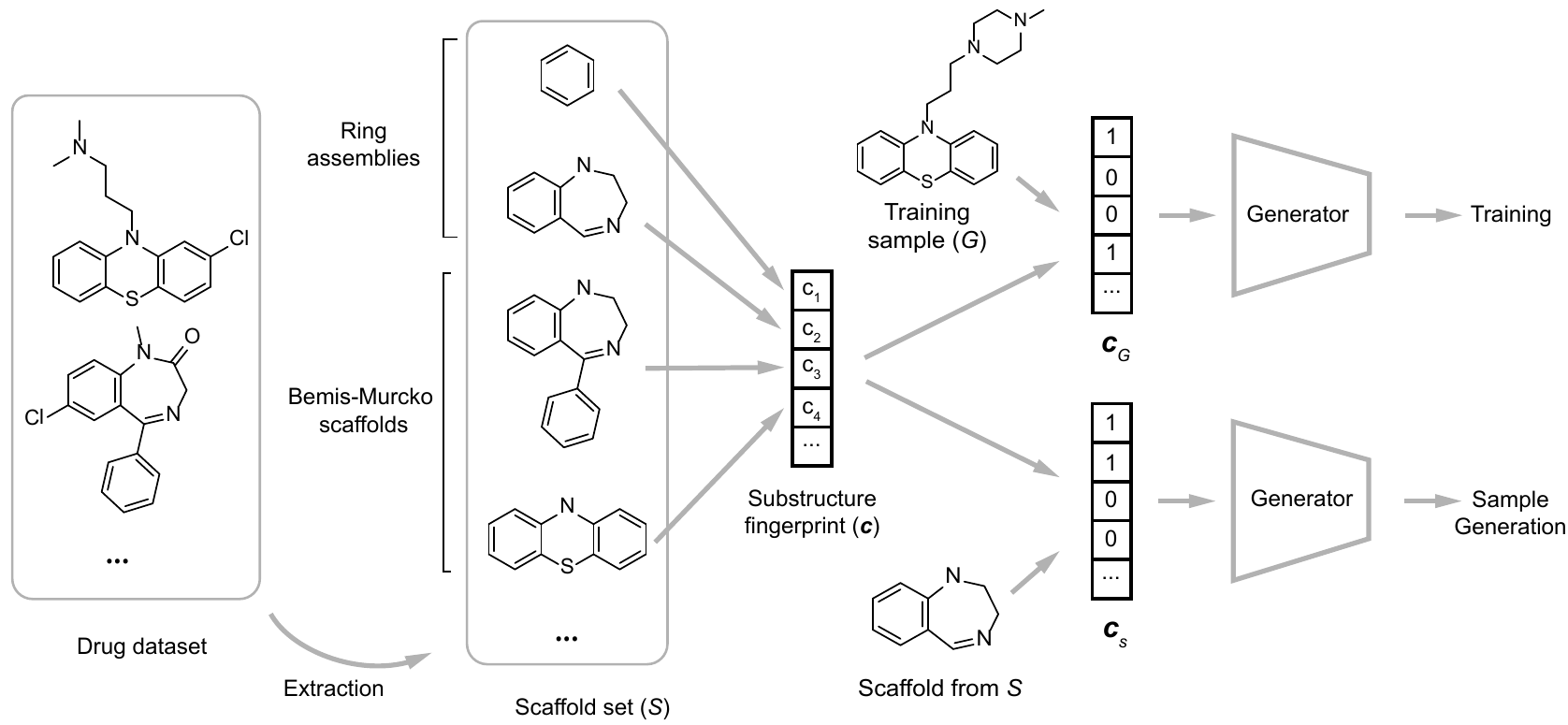}
            \caption{
                \csentence{Workflow for scaffold based molecule generation.}
                Scaffold set $S$ is first extracted from compounds in DrugBank. The conditional code ${\bf c}$ is set to be the 
                substructure fingerprint based on $S$. Training is performed with the training samples labeled with ${\bf c}_G$. After training, scaffold based generation is performed using the fingerprint ${\bf c}_s$ of the query scaffold $s \in S$.
            }
            \label{fig:scaffold_network}
          \end{figure*}

        Here, conditional graph generative model is applied to generate compounds containing scaffold $s$, which is drawnfrom the pre-defined scaffold set $S=\{s_i \}_{i=1}^{N_S}$. The set $S$ is extracted from the list of approved drugs in DrugBank\cite{drugbank}. Two types of structures are extracted from the molecules to construct S: (1) the Bemis-Murcko scaffolds, and (2) ring assemblies. Ring assemblies are included in S since we found that including extra structural information beside Bemis-Murcko scaffolds helps to improve the conditional generation performance. Detailed scaffold extraction workflow is shown in Supplementary Text 2 (Additional file 1). For each molecule $G$, the conditional code ${\bf c}=(c_1, c_2, ..., c_{N_S})$ is set to be the binary vector such that $c_i=1$ if $G$ contains $s_i$ as substructure, and $c_i=0$ otherwise. We refer ${\bf c}$ as the scaffold fingerprint of $G$, since it can in fact be viewed as a substructure fingerprint based on scaffold set $S$. To generate molecule containing substructure $s \in S$, the fingerprint ${\bf c}_s$ for $s$ is used as conditional code. The output should contain two type of molecules:

        \begin{enumerate}
            \item 	Molecules containing $s$ as its Bemis-Murcko scaffold.
    
            \item Molecules whose Bemis-Murcko scaffold contains $s$ but does not reside inside $S$.
    
        \end{enumerate}
    
        The procedure is better demonstrated in Figure \ref{fig:scaffold_network}. Using this method, a detailed control can be performed on the scaffold of the output structure.
    
    \subsection*{Generation Based on Synthetic Accessibility and Drug-likeness}
    
        Drug-likeness and synthetic accessibility are two properties that have significant importance in the development of novo drug candidate. Drug-likeness measures the consistency of a given compound with the currently known drugs in terms of the structural or physical properties and is frequently used to filter out obvious non-drug like compounds in the early phase of screening\cite{review_druglike_1,review_druglike_2}. Synthetic accessibility is also an important property for de novo drug design since subsequent experimental validation requires synthesis of the given compound\cite{sa}. In this task, the model is required to generate molecules according to a given level of drug-likeness and synthetic accessibility. The drug-likeness is measured using the Quantitative Estimate of Drug-likeness (QED)\cite{qed}, and synthetic accessibility is evaluated using the SA score\cite{sa}. The conditional code ${\bf c}$ is defined as ${\bf c}=(QED,SA)$, where the QED and SA score is all calculated using RDKit\cite{rdkit}.
    
        In practice, instead of specifying a single value of QED and SA score, we often use intervals to express the requirements for desired output molecules. This means that we are required to sample molecules from the distribution $p_{\boldsymbol \theta} (G|{\bf c}\in C)=\mathbb{E}_{{\bf c}\sim p({\bf c}|{\bf c}\in C) } [p_{\boldsymbol \theta} (G|{\bf c})]$, where the generation requirement is described as a set $C$ instead of a single point ${\bf c}$. The sampling involves a two-step process by first drawing {\bf c} from $p({\bf c}|{\bf c}\in C)$, and then drawing $G$ from $p_{\boldsymbol \theta} (G|{\bf c})$. Sampling from $p({\bf c}|{\bf c}\in C)$ can be achieved by first sample ${\bf c}$ from $p({\bf c})$ using molecules from the test set, then filter ${\bf c}$ according to the requirement ${\bf c}\in C$.
    
    \subsection*{Designing Dual Inhibitor Against JNK3 and GSK-3$\beta$}
        With the ability to model multiple requirements at once, conditional generative models can be used to design compounds with specific activity profiles for multiple targets. Here, we consider the task of designing dual inhibitors against both c-Jun N-terminal kinase 3 (JNK3) and glycogen synthase kinase-3 beta (GSK-3$\beta$). Both of the two targets are serine/threonine (S/T) kinases, and have shown to be related to the pathogenesis of various types of diseases\cite{review_jnk, review_gsk}. Notably, both JNK3 and GSK-3$\beta$ are shown to be potential target in the treatment of Alzheimer’s disease (AD). Jointly inhibiting JNK3 and GSK-3$\beta$ may provide potential benefit for the treatment of AD.
    
        The conditional code is set to be ${\bf c}=(c_{JNK3},c_{GSK-3\beta})$, where $c_{JNK3}$, $c_{GSK-3\beta}$ are binary values indicating whether the compound is active against JNK3 and GSK-3$\beta$. For compounds in the ChEMBL dataset, $c_{JNK3}$ and $c_{GSK-3\beta}$ are labeled using a separately trained predictor. Random forest (RF) classifier, which has been demonstrated to provide good performance for kinase activity prediction\cite{rf}, is used as the predictor for GSK-3$\beta$ and JNK3 activity, with ECFP6 (extended connectivity fingerprint\cite{ecfp} with a diameter of 6) as the descriptor. The predictive model is trained using activity data from ExCAPE-DB\cite{excape}, which is an integrated database with activity values from ChEMBL and PubChem\cite{pubchem}. Workflow for data extraction and predictor training is provided in Supplementary Text 3. It is found that there is only 1.2\% of molecules in ChEMBL that is predicted to be active against JNK3 or GSK-3$\beta$. This imbalance results in low enrichment rate during conditioned generation. For better result, the model is first trained under the unconditioned setting, and then fine-tuned based on the 1.2\% molecules mentioned above.
        \subsection*{Training Details}
        The graph generative models are trained using the ChEMBL dataset. The data processing workflow largely follows Olivecrona et al \cite{slm_1}, as described in Supplementary Text 1. MXNet\cite{mxnet} is used to implement the networks, and Adam optimizer\cite{adam} is used for network training. An initial learning rate of 0.001 is used together with a decay rate of 0.001 for every 100 iterations. Other parameters of the optimizer are set to be the default values suggested in \cite{adam} (that is, $\beta_1=0.9, \beta_2=0.999$ and $\epsilon=10^{-8}$). The training lasts for 5 epochs, and the size of each mini-batch is set to 200 during the training.
    
        During training, the decoding route is drawn from the distribution $q_\alpha (r|G)$. We tried three $\alpha$ values: 1.0, 0.8 and 0.6, as discussed previously. For  $\alpha = 1.0$, $k$ is set to 1 and the training can be performed on a single Nvidia GeForce GTX 1080Ti GPU for both MolMP and MolRNN. The training lasts for 14h for MolMP and 16h for MolRNN. For $\alpha=0.8$ and $\alpha=0.6$, $k$ is set to 5 and the training is performed synchronously on 4 GPUs. The training lasts for 30h for MolMP and 35h for MolRNN.
    
        For scaffold based and property based generation tasks, the conditonal graph generator is trained using the same setting as unconditional model. For the generation of GSK-3$\beta$ and JNK3 inhibitors, the model is first trained using the full dataset, and the fine tuned on the subset that is predicted to be active against GSK-3$\beta$ or JNK3. The fine-tuning uses a learning rate of 0.0001 and a decay rate of 0.002 for every 100 iterations. The fine-tuning lasts for 10 epochs, and takes 1h to finish.
    
        In theory, the hyperparameters for the models mentioned above, including the training condition (batch size, learning rate, decay rate, $\beta_1$, $\beta_2$), model architectures(the number of convolutional layers, the hidden size in each layer) as well as $\alpha$, should be optimized to achieve the best performance. However, due to the computational cost of both MolMP and MolRNN, we are unable to systematically optimize the hyperparameters. A througout discussion is only given for $\alpha$, which determines the degree of randomness of $q_\alpha$. No optimization is performed on model architecture except fitting it into the memory. 
    
    \subsection*{SMILES Based Methods}
    
        The proposed graph-based model is compared with several SMILES based models for model performance and sample quality. Two type of methods, variational autoencoder (VAE) and language model (LM), are considered in this comparison. The implementation of SMILES VAE follows Gómez-Bombarelli et al\cite{svae_1}. The encoder contains three 1D convolutional layers, with 9, 9, 10 filters and 9, 9, 11 kernels each, and a fully connected layer with 435 hidden units. The model uses 196 latent variables and a decoder with three GRU layers with 488  hidden units. VAE for sequential data faces from the issue of “optimization challenge”\cite{rnn_vae, vlae}.  While the original implementation uses KL-annealing to tackle this problem, we follow the method provided  by Kingma et al\cite{iaf} by controlling the level of free bits. This offers higher flexibility and stability  compared with KL-annealing. We restrict the minimal level of free bits to 0.03 for each latent variable.
        
        For LM, two types recurrent units are adopted. The first type uses GRU, and includes two architectures: the first architecture (SMILES GRU1) consists of three GRU layers with 512 hidden units each, and the second (SMILES GRU2), uses a wider GRU architecture with 1024 units, following the implementation by Olivecrona et al\cite{slm_1}. Beside GRU, we also included a LSTM based SMILES language model following Segler et al\cite{slm_3}. This architecture uses three LSTM layers, each with 1024 units.
    
    \subsection*{Evaluation Metrics}
    
        Several metrics have been employed to evaluate the performance of generative models:
        
        \subsubsection*{Sample Validity}
            To test whether the generative models are capable of producing chemically correct outputs, 300,000 structures are generated for each model, and subsequently evalulated by RDKit for the rate of valid outputs. We also evaluate the ability of each model to produce novel structures. This is done by accessing the rate of generated compounds that do not occure inside the training set.
    
        \subsubsection*{$D_{KL}$ and $D_{JS}$ for Molecular Properties}
            A good molecule generator should correctly model the distribution of important molecular properties. Therefore, the distribution of molecular weight (MW), log-partition coefficient (LogP) and QED between the generated dataset ($p_g$) and the test set ($p_{data}$) is compared for each method, using Kullback–Leibler divergence ($D_{KL}$):
            \begin{equation}
                D_{KL}(p_g||p_{data}) = \int_{\mathbb{R}}{p_g(x)\log{\frac{p_g(x)}{p_{data}(x)}}dx}
            \end{equation}
            and Jensen–Shannon  divergence($D_{JS}$):
            \begin{equation}
                \begin{split}
                    D_{JS}(p_g||p_{data}) = & \frac{1}{2}D_{KL}(p_g||\frac{p_g +p_{data}}{2}) + \\
                    & \frac{1}{2}D_{KL}(p_{data}||\frac{p_g + p_{data}}{2})
                \end{split}
            \end{equation}
            $D_{KL}$ and $D_{JS}$ are widely used in deep generated models for both training \cite{vae, gan} and evaluation \cite{gan_eval}. Here, the two values are determined using kernel density method implemented in SciPy \cite{scipy}. We used a gaussian kernel with bandwidth selected based on Scott’s Rule\cite{scott}.

        \subsubsection*{Negative Log-Likelihood}
            The model performance is also evaluated using the negative log-likelihood (NLL) on the test set $\{G_i \}_{i=1}^N$. To offer comparison between graph and SMILES based generative model, NLL is evaluated using the canonical ordering as follows:
    
            \begin{equation}
                NLL = -\frac{1}{N}\sum_{i=1}^N{\log{p_{\boldsymbol \theta}(G_i,r_i^*)}}
            \end{equation}
    
            Note that for graph based models, NLL is only reported for models trained on $\alpha = 1$. For models using $\alpha < 1$, the value caluclated above can not be directly compared between different models. Therefore, we rely more on other metrics such as $D_{KL}$ and $D_{JS}$. Also, for SMILES VAE, importance sampling is performed to obtain a tighter bound. The number of samples is set to be 100 ($k=100$).
    
        \subsubsection*{Performance Metrics for Conditional Generative Models}
            For discrete conditional codes ${\bf c}$, let $M_{\bf c}$ be the set containing molecules sampled from distribution 
            $p_{\boldsymbol \theta} (G|{\bf c})$. $M_{\bf c}$ is obtained by first sampling molecule graphs conditioned on ${\bf c}$ 
            and then removing invalid molecules. The size of $|M_{\bf c}|$ is set to 1,000. Let $N_{{\bf c}{\bf c}'}$ be the set of molecules in $M_{\bf c}$ that satisfy the condition ${\bf c}'$ (${\bf c}'$ may be different from ${\bf c}$). The ratio $K_{{\bf c}{\bf c}'}$ is defined as:
    
            \begin{equation}
                K_{{\bf c}{\bf c}'} = \frac{|N_{{\bf c}{\bf c}'}|}{|M_{\bf c}|}
            \end{equation}
    
            The matrix $K_{{\bf c}{\bf c}'}$ can be used to evaluate the ability of the model to control the output based on  conditional code ${\bf c}$. When ${\bf c}={\bf c}'$, this value gives the rate of correctly generated outputs, denoted  by $R_{\bf c}$. High quality conditional models should have a high value of $R_{\bf c}$ and low values of  $K_{{\bf c}{\bf c}'}$ for ${\bf c} \ne {\bf c}'$. In paractice, we find that the value of $K_{{\bf c}{\bf c}'}$ for scaffold and property based generation is significantly samller than $R_{\bf c}$ and have relatively low influence on the model's performance. Therefore, the result of $K_{{\bf c}{\bf c}'}$ is omitted for scaffold and property based task, and is only reported for the task of kinase inhibitor design.
    
            Let $R_{\bf c}^0$ be the rate of molecules in the training data that satisfy condition ${\bf c}$. The enrichment over random $EOR_{\bf c}$ is defined as:
    
            \begin{equation}
                EOR_{\bf c} = \frac{R_{\bf c}}{R_{\bf c}^0}
            \end{equation}
    
            The definition is similar to that used in previous work\cite{slm_3}, except that in their implementation $R_{\bf c}^0$  is calculated using the generated samples from the unconditioned model $p_{\boldsymbol \theta} (G)$. For continuous codes, a subset $C$ of the conditional code space is used to describe the generation requirements. $M_C$  is sampled from $p_{\boldsymbol \theta} (G|{\bf c}\in C)$, and values for $K_{CC'}$, $R_C$ and $EOR_C$ can be  calculated in a similar manner.
    
            For target based generation task, the rate of reproduced molecules is also reported following previous  works\cite{slm_1, slm_3}. Take JNK3 as an example. During the evaluation, two sets of outputs are  generated using two conditions: JNK3(+), GSK-3$\beta$ (-) and JNK3(+), GSK-3$\beta$(+).  The two set of outputs are denoted $M_{{\bf c}_1}$and $M_{{\bf c}_2}$respectively. Here, the size of $|M_{{\bf c}_1}|$  and $|M_{{\bf c}_2}|$ are both set to 50,000. Let $T$ be the set containing the active molecules within the test set  of JNK3. The rate of reproduced molecules ($reprod$) is calculated as:
    
            \begin{equation}
                reprod = \frac{|(M_{{\bf c}_1} \cup M_{{\bf c}_2})\cap T|}{|T|}
            \end{equation}
    
            For GSK-3$\beta$, the calculation can be done in a similar manner.
            
            Finally, we access the diversity of the generated outputs by conditional models using the internal diversity $I$ proposed in \cite{chem_gan}:
    
            \begin{equation}
                I(M) = \frac{1}{|M|^2}\sum_{(x, y)\in M\times M}T_d(x, y)
            \end{equation}
    
            Where $M$ is the set of sampled molecules, and $T_d(x, y)$ is the Tanimoto-distance between the two molecules $x$ and $y$. $T_d(x, y)$ is defined using the Tanimoto-similarity $T_s$: $T_d(x, y) = 1 - T_s(x, y)$.

  \section*{Results and Discussion}
    \subsection*{Model Performance and Sample Quality}

      \label{model-performance}

      \begin{figure}[b!]
        \includegraphics{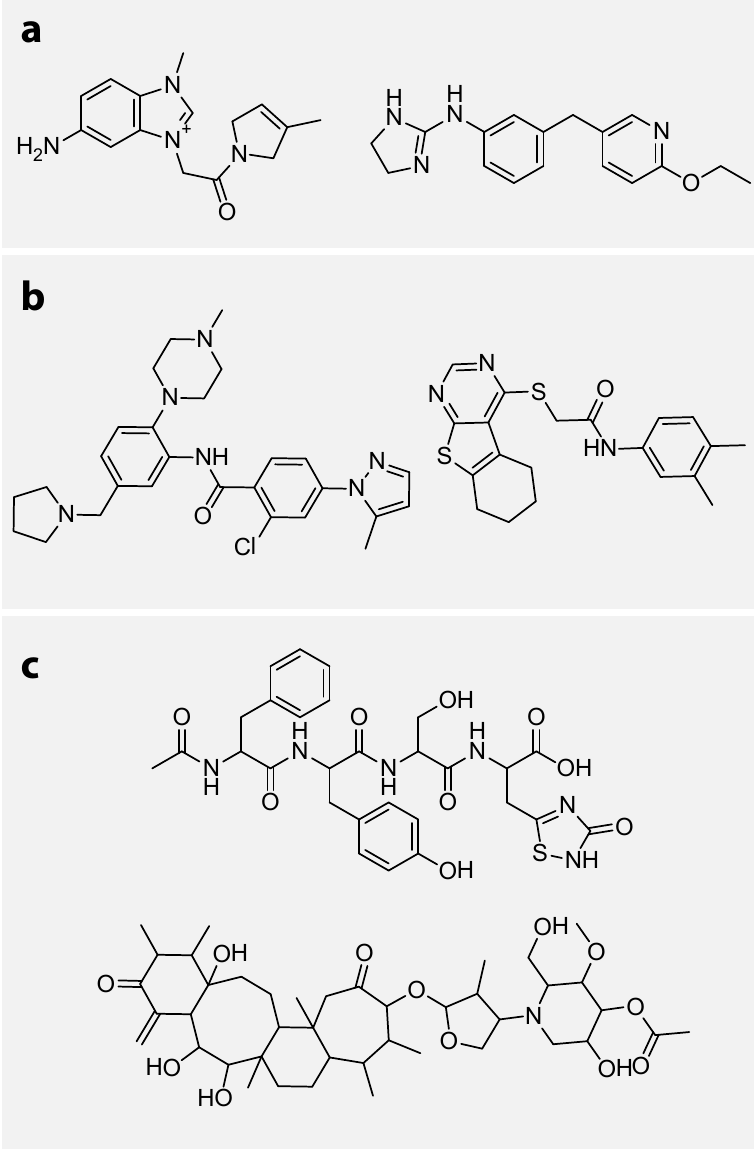}
        \caption{
            \csentence{Output samples by MolRNN}
            The outputs are grouped by molecular weight(MW): {\bf a}. MW\textless 300; {\bf b}. 300$\leq$MW\textless 500; {\bf c}. MW$\geq$500
        }
        \label{fig:samples}
      \end{figure}

      Several randomly generated samples from MolRNN are grouped by molecular weight and shown in Figure \ref{fig:samples}. The comparison between SMILES based and graph based models (MolMP and MolRNN) have been performed, and the results is summarized in Table \ref{tab:performance_1} and Table \ref{tab:performance_2}. We first analysed the model performance in terms of NLL. According to the result, MolRNN is able to achieve the best performance with $NLL=24.08$. As for MolMP, although it is unable to outperform SMILES GRU2 and SMILES LSTM, it achieves better performance compared with SMILES GRU1 and SMILES VAE. It should be noted that SMILES GRU1 contains $4 \times 10^6$ parameters, while MolMP only contains $1\times10^6$. This indicates that graph based models are more efficient in parameter usage. It should also be noted that the NLL values used in this comparison are only relatively  loose bonds as it is evaluated using only deterministic decoding route. Therefore, we focuces more on other evaluation metrics that are discussed below.

      \begin{figure}[b!]
        \includegraphics{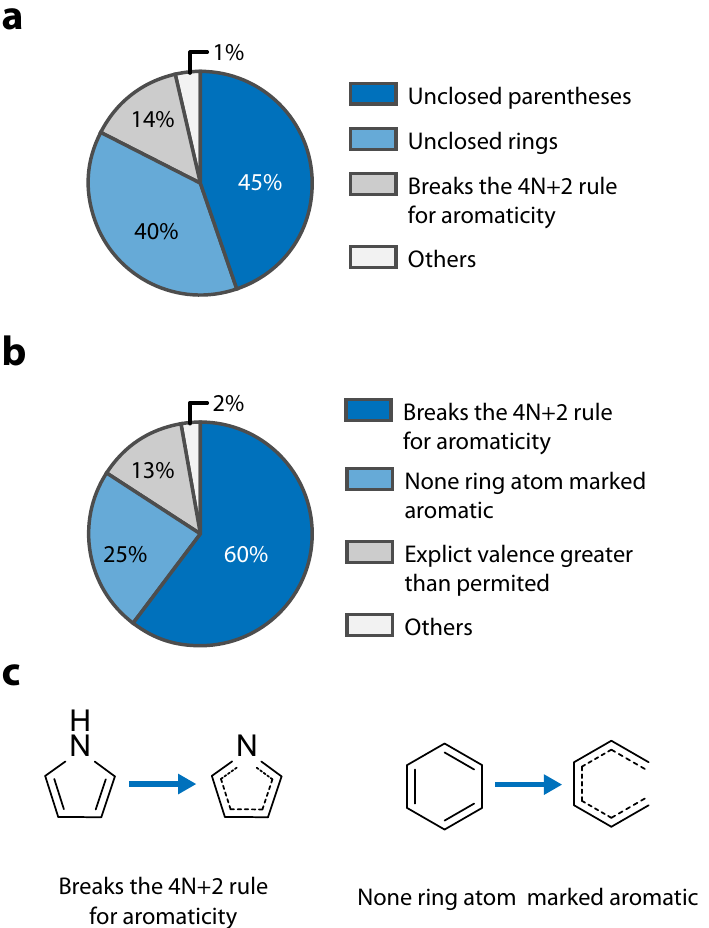}
        \caption{
            Common mistakes made by: {\bf a}. SMILES based model and {\bf b} graph based model. {\bf c} Examples of broken aromaticity occurred during graph generation.
        }
        \label{fig:error}
      \end{figure}

      \begin{table*}
        \caption{Comparison between SMILES based and graph-based generators in NLL and output validity. Results are reported as $Mean \pm StdDev$. The model giving the best performance in each metric is highlighted in boldface}
        \begin{tabular}{lcccc}
            \toprule
            Model & NLL & \% valid & \% novel & \% valid \& novel \\
            \midrule
            SMILES VAE &$30.39\pm 0.25$ & $0.804\pm 0.016$ & ${\bf 0.986\pm 0.000}$ & $0.793\pm 0.016$ \\
            SMILES GRU1&$27.57\pm 0.03$ & $0.886\pm 0.002$ & $0.984\pm 0.000$ & $0.872\pm 0.002$ \\
            SMILES GRU2&$24.45\pm 0.02$ & $0.932\pm 0.002$ & $0.965\pm 0.001$ & $0.899\pm 0.002$ \\
            SMILES LSTM&$25.43\pm 0.04$ & $0.935\pm 0.006$ & $0.975\pm 0.001$ & $0.912\pm 0.006$ \\
            \midrule
            MolMP ($\alpha=1.0$)& $26.25\pm 0.02$ & $0.952\pm 0.002$ & $0.98\pm 0.001$ & $0.933\pm 0.001$ \\
            MolMP ($\alpha=0.8$)&- & $0.962\pm 0.002$ & $0.984\pm 0.001$ & $0.946\pm 0.001$ \\ 
            MolMP ($\alpha=0.6$)&- & $0.963\pm 0.001$ & $0.988\pm 0.001$ & $0.951\pm 0.001$ \\
            \midrule
            MolRNN ($\alpha=1.0$)&${\bf 24.08\pm 0.03}$ & $0.967\pm 0.001$ & $0.959\pm 0.000$ & $0.928\pm 0.001$ \\
            MolRNN ($\alpha=0.8$)&- & ${\bf 0.970\pm 0.001}$ & $0.976\pm 0.001$ & $0.947\pm 0.001$ \\
            MolRNN ($\alpha=0.6$)&-& ${\bf 0.970\pm 0.001}$ & $0.985\pm 0.000$ & ${\bf 0.955\pm 0.001}$ \\
            \bottomrule
        \end{tabular}
        \label{tab:performance_1}
      \end{table*}

      \begin{table*}
        \caption{Comparison between SMILES based and graph-based generators in $D_{KL}$($\times 10^{-3}$) and $D_{JS}$($\times 10^{-3}$). Results are reported as $Mean \pm StdDev$. The model giving the best performance in each metric is highlighted in boldface}
        \begin{tabular}{lcccccc}
            \toprule
            & \multicolumn{2}{c}{MW} & \multicolumn{2}{c}{LogP} & \multicolumn{2}{c}{QED} \\
            \cmidrule{2-7}
            Model & $D_{KL}$ & $D_{JS}$ & $D_{KL}$ & $D_{JS}$ & $D_{KL}$ & $D_{JS}$ \\
            \midrule
            SMILES VAE & $13.5\pm 0.6$ &$3.6\pm 0.2$ &$3.9\pm 0.4$ &$0.9\pm 0.1$ &$2.6\pm 0.4$ &$0.6\pm 0.1$ \\
            SMILES GRU1 & $8.6\pm 0.4$ &$2.3\pm 0.1$ &$3.1\pm 0.3$ &$0.7\pm 0.0$ &$1.5\pm 0.3$ &$0.3\pm 0.1$ \\
            SMILES GRU2 & $7.8\pm 0.3$ &$2.0\pm 0.1$ &${\bf 1.4\pm 0.2}$ &${\bf 0.3\pm 0.0}$ &$2.2\pm 0.3$ &$0.5\pm 0.1$ \\
            SMILES LSTM & $6.5\pm 0.7$ &$1.8\pm 0.2$ &$3.4\pm 1.2$ &$0.8\pm 0.3$ &$1.9\pm 1.3$ &$0.4\pm 0.3$ \\
            \midrule
            MolMP ($\alpha=1.0$) & $11.5\pm 1.3$ &$3.4\pm 0.4$ &$7.0\pm 1.8$ &$1.7\pm 0.4$ &$5.3\pm 1.2$ &$1.3\pm 0.3$ \\
            MolMP ($\alpha=0.8$) & $8.3\pm 1.6$ &$2.4\pm 0.5$ &$4.3\pm 1.2$ &$0.9\pm 0.2$ &$2.7\pm 0.8$ &$0.6\pm 0.2$ \\
            MolMP ($\alpha=0.6$) & $8.4\pm 1.0$ &$2.4\pm 0.3$ &$5.0\pm 1.3$ &$1.1\pm 0.4$ &$3.0 \pm 0.9$ &$0.7\pm 0.2$ \\
            \midrule
            MolRNN ($\alpha=1.0$) & $5.0\pm 0.6$ &$1.4\pm 0.2$ &$2.8\pm 0.5$ &$0.7\pm 0.1$ &$2.0\pm 0.6$ &$0.5\pm 0.1$ \\
            MolRNN ($\alpha=0.8$) & $4.1\pm 0.7$ &$1.1\pm 0.2$ &$1.6\pm 0.3$ &$0.3\pm 0.1$ &${\bf 1.0\pm 0.2}$ &${\bf 0.2\pm 0.0}$ \\
            MolRNN ($\alpha=0.6$) & ${\bf 3.3\pm 0.2}$ &${\bf 0.9\pm 0.1}$ &$3.0\pm 0.4$ &$0.5\pm 0.1$ &$1.1\pm 0.4$ &$0.2\pm 0.1$ \\
            \bottomrule
        \end{tabular}
        \label{tab:performance_2}
      \end{table*}

      In terms of the rate of valid outputs and the rate of valid and novel outputs, both MolRNN and MolMP outperform all SMILES based methods. It is also noted that changing $\alpha$ from 1.0 to 0.8 can significantly increase the rate of valid outputs for both MolMP and MolRNN. Further decreasing $\alpha$ can produce only margincal effect. The high validity in output structures of graph-based model is not surprising as the generation of SMILES poses much stricter rules to the output compared with the generation of molecular graphs. 

      Figure \ref{fig:error}{\bf a} and Figure \ref{fig:error}{\bf b} summarize respectively the common mistakes made by SMILES-based and graph-based model during generation. Results in Figure \ref{fig:error}{\bf a} show that the most common cause of invalid output for SMILES based models is grammar mistakes, such as unclosed parentheses or unpaired ring numberings. But for the graph-based model, the majority of invalid output is caused by broken aromaticity, as demonstrated in Figure \ref{fig:error}{\bf c}. This is likely a result of stepwise decoding pattern of graph-based models, as the decoder can only see part of the aromatic structure during generation, while  the determination of aromaticity requires the information of the entire ring. It is also observed that mistakes  related to atom valance are relatively minor, meaning that those rules are easy to learning using graph convolution.  

      Graph-based methods also have the advantage of giving the highly interpretable outputs compared with SMILES.  This means that a large portion of invalid outputs can be easily corrected if necessary. For example,  broken aromaticity can be restored by literately refining the number explicit hydrogens of aromatic atoms,  and unclosed aromatic rings can be corrected simply by connecting the two ends using a new aromatic bond.  Though possible, those corrections may introduce additional bias to the output samples depending on the  implementation, thus not adopted in the subsequent evaluations.

      Next, we investigate the ability for the generators to learn the distribution of molecular properties, as demonstrated in Table \ref{tab:performance_2}. Results have shown that MolRNN gives the best performance in $D_{KL}$ and $D_{JS}$ for molecular weight (MW) and QED, while SMILES GRU2 gives the best performance for LogP. For MolMP, although it is able to outperform SMILES GRU1 in NLL, it fails to give better performance in $D_{KL}$ and $D_{JS}$. This observation suggest that the molecule level recurrent unit in MolRNN can significantly imporved the ability for the model to learn information about the data distribution.

      \begin{figure*}[t!]
        \includegraphics{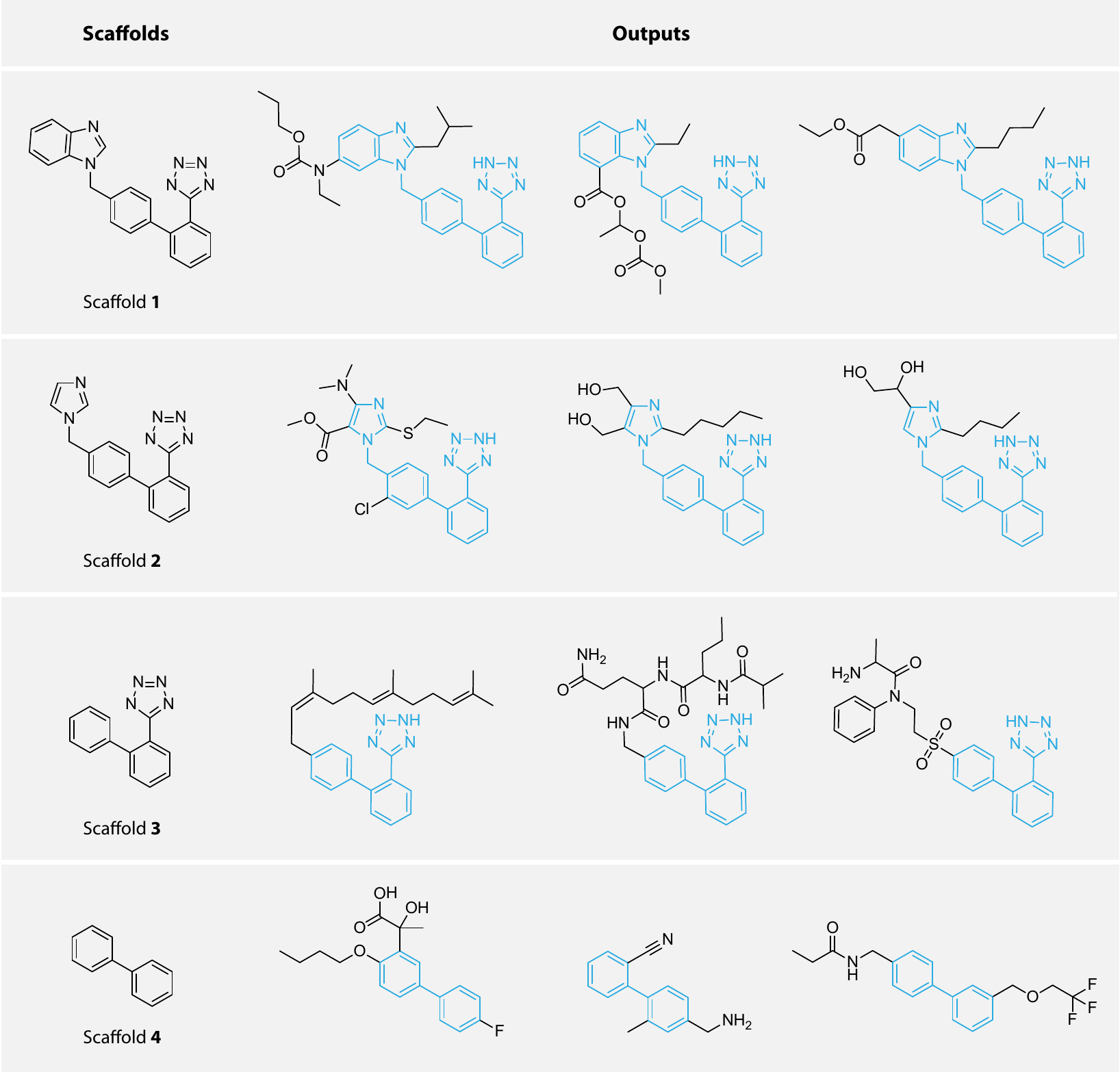}
        \caption{
            \csentence{Results of scaffold based molecule generation using scaffold {\bf 1-4} as conditions}
        }
        \label{fig:scaffold_sample}
      \end{figure*}

      When it comes to the influence of $\alpha$ to $D_{KL}$ and $D_{JS}$, it is found that changing $\alpha$ from 1.0 to 0.8 can significantly improve the perforamnce of MolMP and MolRNN for all molecular properties. Further decreasing $\alpha$ to 0.6 will have different effect for MolMP and MolRNN. For MolMP, this will hurt the overall performance of $D_{KL}$ and $D_{JS}$, while for MolRNN, this will inprove the performance for molecular weight, but will significantly decrease the performance of LogP. Overall, $\alpha=0.8$ will be a better choise for MolMP, and $\alpha=0.6$ will be more suited for MolRNN.

      Generally, MolRNN have showed significant advantages among all generative mdoels considered. In the subsequent evaluation of conditonal generative models, the best performing graph based model (MolRNN) and the best performing SMILES based model (SMILES GRU2) are implemented as conditonal models and are compared among all tasks.
    \subsection*{Scaffold-Based Generation}
      \label{sec:scaffold}


      In the first task, conditional generative models are trained to produce molecules based on a given scaffold. To illustrate the result, scaffold {\bf 1}, extracted from the antihypertensive drug Candesartan 
      , is used as an example, along with several related scaffolds (scaffold {\bf 2-4}) derived from scaffold {\bf 1} (Figure \ref{fig:scaffold_sample}). Conditional codes ${\bf c}$ are constructed for each type of scaffold, and output structures are produced according to the corresponding code.

      \begin{table*}
        \caption{Performance of graph based and SMILES based model on scaffold diversification tasks. Results are reported as $Mean \pm StdDev$. The model giving the best performance in each metric is highlighted in boldface}
        \begin{tabular}{ccccccc}
            \toprule
            Condition ($\bf c$) & $R_0$ & Model & \% valid & $R_{\bf c}$ & $EOR_{\bf c}$ & Diversity \\
            \midrule
            \multirow{2}{*}{scaffold {\bf 1}} & \multirow{2}{*}{$7.9\times 10^{-5}$} & Graph &${\bf 0.931\pm 0.008}$ & $0.86\pm 0.03$ & 10865 & $ 0.496\pm 0.015$ \\ 
            &&SMILES&$0.924\pm 0.005$ & ${\bf 0.87\pm 0.01}$ & {\bf 10976} & ${\bf 0.498\pm 0.015}$ \\ 
            \cmidrule{3-7}
            \multirow{2}{*}{scaffold {\bf 2}} & \multirow{2}{*}{$1.1\times 10^{-4}$} & Graph &${\bf 0.900\pm 0.016}$ & $0.77\pm 0.04$ & 6972 & ${\bf 0.531\pm 0.02}$ \\ 
            &&SMILES&$0.896\pm 0.011$ & ${\bf 0.84\pm 0.01}$ &{\bf  7607} & $0.495\pm 0.015$ \\ 
            \cmidrule{3-7}
            \multirow{2}{*}{scaffold {\bf 3}} & \multirow{2}{*}{$7.9\times 10^{-5}$} & Graph &${\bf 0.940\pm 0.019}$ & ${\bf 0.56\pm 0.08}$ & {\bf 7086} & $0.683\pm 0.023$ \\
            &&SMILES&$0.898\pm 0.024$ & $0.37\pm 0.07$ & 4623 & ${\bf 0.704\pm 0.022}$ \\ 
            \cmidrule{3-7}
            \multirow{2}{*}{scaffold {\bf 4}} & \multirow{2}{*}{$5.8\times 10^{-3}$} & Graph &${\bf 0.982\pm 0.001}$ & ${\bf 0.88\pm 0.01}$ & {\bf 151} & $0.815\pm 0.001$ \\ 
            &&SMILES&$0.969\pm 0.002$ & ${\bf 0.88\pm 0.00}$ & {\bf 151} & ${\bf 0.823\pm 0.00}$ \\ 
            \bottomrule
        \end{tabular}
        \label{tab:scaffold_result}
      \end{table*}
      
      Results for both the SMILES based and graph based conditional generator are given in Table \ref{tab:scaffold_result}. In terms of output validity, graph based model is able to produce a higher fraction of valid outputs for scaffolds {\bf 1-4}, compared with SMILES based methods. This is similar to the results of unconditional models
      
      In terms of the rate of correctly generated outputs ($R_{\bf c}$), although the  models are unable to achieve 100\% correctness, the $R_{\bf c}$ results are significantly higher than $R_{\bf c}^0$, offering high enrichment rate over random. Both graph based and SMILES based model are able to achieve $EOR_{\bf c} > 1,000$ for scaffold {\bf 1-3} as well as $EOR_{\bf c} > 100$ for scaffold {\bf 4}, showing promising ability for the model to produce enriched output according to the given scaffold query. By comparing the result of $R_{\bf c}$ between the two type of architectures, it is found that graph based model have a higher performance for scaffold {\bf 3}, while SMILES based method have a higher performance for scaffold {\bf 2}. The two model have similar performance for scaffold {\bf 1} and scaffold {\bf 4}. 

      The structural diversity of the output samples is also evaluated for each model. It is found that SMILES based model tends to produce outputs that are more diverse compared with graph based model, except for scaffold {\bf 3}. This may indicate that the graph based model tends to be slightly overtrained compared with SMILES based model. However, those differences are relatively minor compared with the standard deviation of each value. 

      \begin{figure}[b!]
        \includegraphics{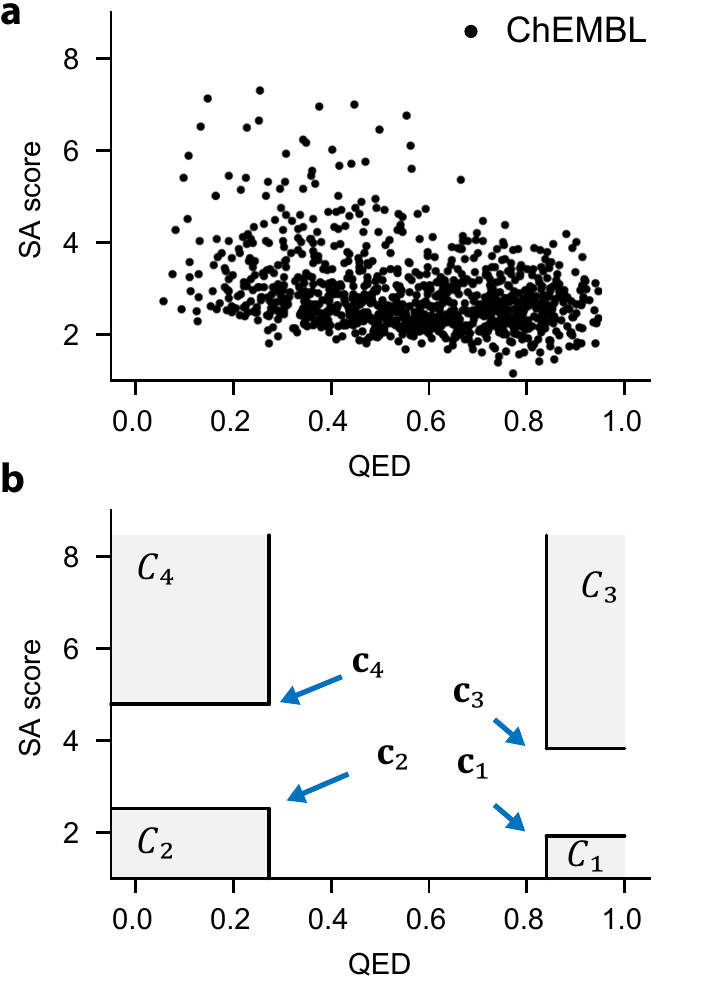}
        \caption{
            \csentence{Location of $C_1\sim C_4$ and ${\bf c_1}\sim {\bf c_4}$:}
            {\bf a}. Distribution of QED and SAscore in the ChEMBL dataset; {\bf b}. Location of the conditional codes ${\bf c}_1$ , ${\bf c}_2$, ${\bf c}_3$ and ${\bf c}_4$ and conditional sets $C_1$, $C_2$, $C_3$ and $C_4$  used in the evaluation.
        }
        \label{fig:prop_conditions}
      \end{figure}

      \begin{table*}
        \caption{Performance of graph based and SMILES based model on property based generation tasks. Results are reported as $Mean \pm StdDev$. The model giving the best performance in each metric is highlighted in boldface}
        \begin{tabular}{ccccccc}
            \toprule
            Condition ($C$) & $R_0$ & Model & \% valid & $R_C$ & $EOR_C$ & Diversity \\
            \midrule
            \multirow{2}{*}{$C_1$} & \multirow{2}{*}{$0.009$} & Graph & ${\bf 0.997\pm 0.000}$ & ${\bf 0.55\pm 0.01}$ & {\bf 61} & $0.814\pm 0.002$ \\ 
            &&SMILES&$0.995\pm 0.001$ & $0.51\pm 0.00$ & 57 & ${\bf 0.827\pm 0.000}$ \\ 
            \cmidrule{3-7}
            \multirow{2}{*}{$C_2$} & \multirow{2}{*}{$0.012$} &Graph &${\bf 0.970\pm 0.002}$ &$ {\bf 0.55\pm 0.01}$ & {\bf 46} & $0.848\pm 0.001$ \\ 
            &&SMILES&$0.944\pm 0.001$ & $0.52\pm 0.00$ & 43 & ${\bf 0.849\pm 0.001}$ \\ 
            \cmidrule{3-7}
            \multirow{2}{*}{$C_3$} & \multirow{2}{*}{$0.011$} &Graph &${\bf 0.957\pm 0.001}$ & ${\bf 0.35\pm 0.01}$ & {\bf 32} & $0.872\pm 0.001$ \\ 
            &&SMILES&$0.894\pm 0.007$ & $0.31\pm 0.00$ & 28 & ${\bf 0.878\pm 0.00}$ \\ 
            \cmidrule{3-7}
            \multirow{2}{*}{$C_4$} & \multirow{2}{*}{$0.008$} &Graph &${\bf 0.929\pm 0.003}$ & ${\bf 0.73\pm 0.01}$ & {\bf 91} & $0.865\pm 0.000$ \\ 
            &&SMILES&$0.613\pm 0.015$ & $0.66\pm 0.00$ & 82 & ${\bf 0.867\pm 0.00}$ \\ 
            \bottomrule
        \end{tabular}
        \label{tab:prop_result}
      \end{table*}

      Several generated samples by graph based model are given for each scaffold in Figure \ref{fig:scaffold_sample}. Recall that the outputs given scaffold $s$ should contain two type of molecules: (1) molecules with $s$ as its Bemis-Murcko scaffold and (2) molecule whose Bemis-Murcko  scaffold contains $s$ but does not reside inside $S$. Both types are observed for scaffold {\bf 1-4} as shown in Figure \ref{fig:scaffold_sample}.  By further investigating the generated samples, it is observed that the model seems to have learnt about the side chains characteristics each scaffold. For example, samples generated from scaffold {\bf 1-3} usually have their  substitutions occur at restricted positions, and frequently contains a long aliphatic side chain. Interestingly,  this actually reflects the structural activity relationship (SAR) for angiotensin II (Ang II) receptor  antagonists\cite{ang_1}. In fact, scaffold {\bf 1-3} have long been treated as a privileged structure against Ang II  receptors\cite{review_scaffold}, and as a result, molecules with scaffold {\bf 1-3} are largely biased to those who matches the SAR rules for the target. When trained with the biased dataset, the model can memorize the underlying structural activity relationship as a byproduct of scaffold based learning. This characteristic is beneficial for the generation of libraries containing specified privileged structures.
      
      \begin{figure*}[t!]
        \includegraphics{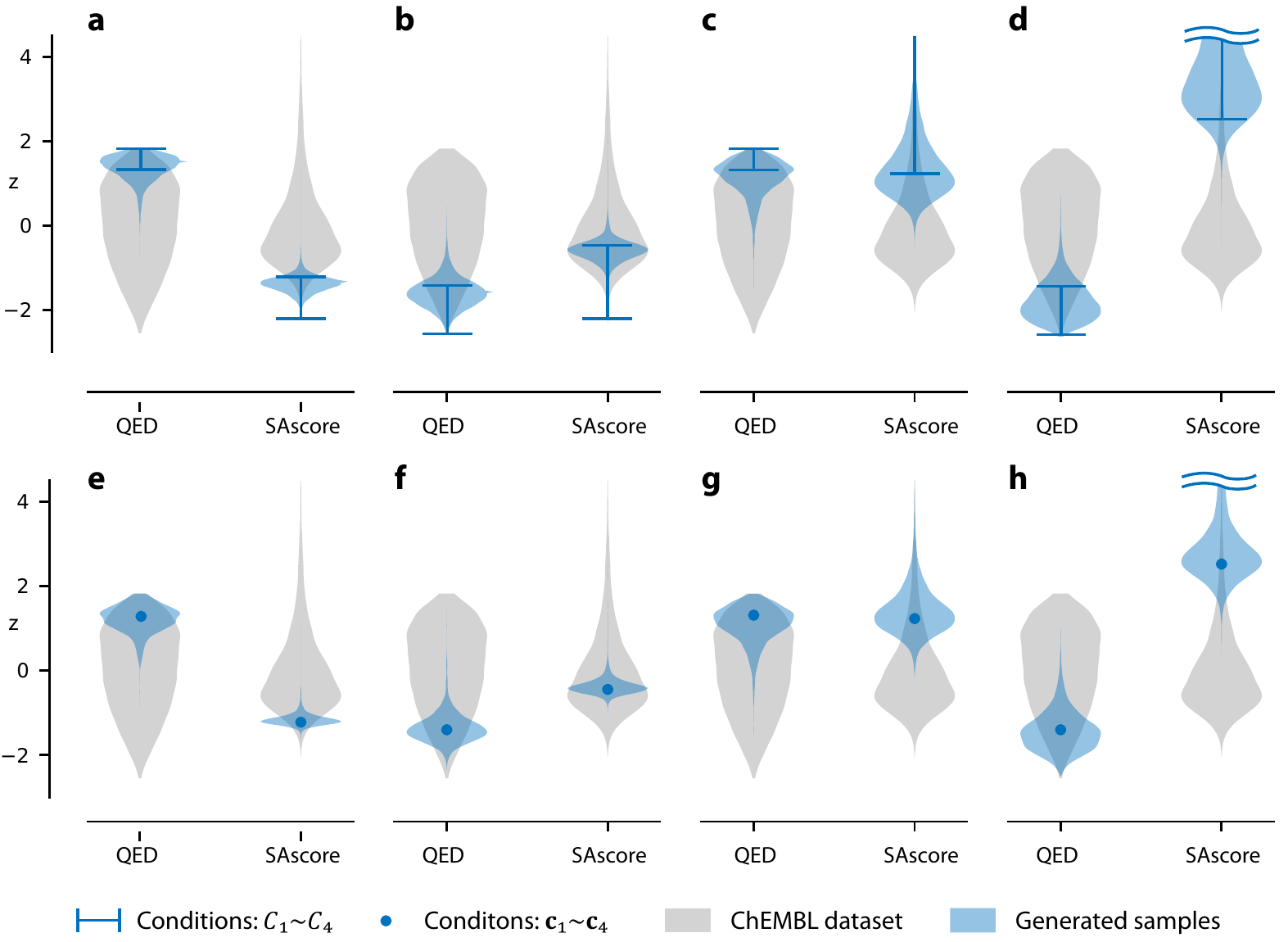}
        \caption{
            \csentence{Distribution of QED and SAscore for generated results:} {\bf a}-{\bf d}: Distribution of QED and SAscore of molecules generated under conditions $C_1$, $C_2$ , $C_3$ , $C_4$ respectively. The conditions $C_1\sim C_4$ are shown as intervals represented by error bar. {\bf e}-{\bf h}: Distribution of QED and SAscore of molecules generated using single point conditions, which are ${\bf c}_1$, ${\bf c}_2$, ${\bf c}_3$ and ${\bf c}_4$ respectively. The conditions ${\bf c_1}\sim {\bf c_4}$ are represented as dots in the plot.
        }
        \label{fig:prop_distributions}
      \end{figure*}

    \subsection*{Generation Based on Drug-likeness and Synthetic Accessibility} 
      In this task, the generative model is used to produce molecules according to the requirement on drug-likeness and  synthetic accessibility. The conditional code is specified as ${\bf c}=(QED,SA)$.  In the first experiments, the models are required to generate molecules based on the following requirements expressed as subsets of conditional code space:
      $C_1=(0.84,1)\times(0,1.9)$, 
      $C_2=(0,0.27)\times(0,2.5)$, 
      $C_3=(0.84,1)\times(3.4,+\infty)$ and $C_4=(0,0.27)\times(4.8,+\infty)$.

      The values are determined from the distribution of QED and SA in ChEMBL dataset (see Figure \ref{fig:prop_conditions}{\bf a}) using the 90\% and 10\% quantile. The conditions are illustrated in Figure \ref{fig:prop_conditions}{\bf d}. The four sets represent four classes of molecules respectively and the first class $C_1$, which contains structures with high drug-likeness and high synthetic accessibility, defines the set of compounds that are most important for drug design.

      \begin{figure*}
        \includegraphics{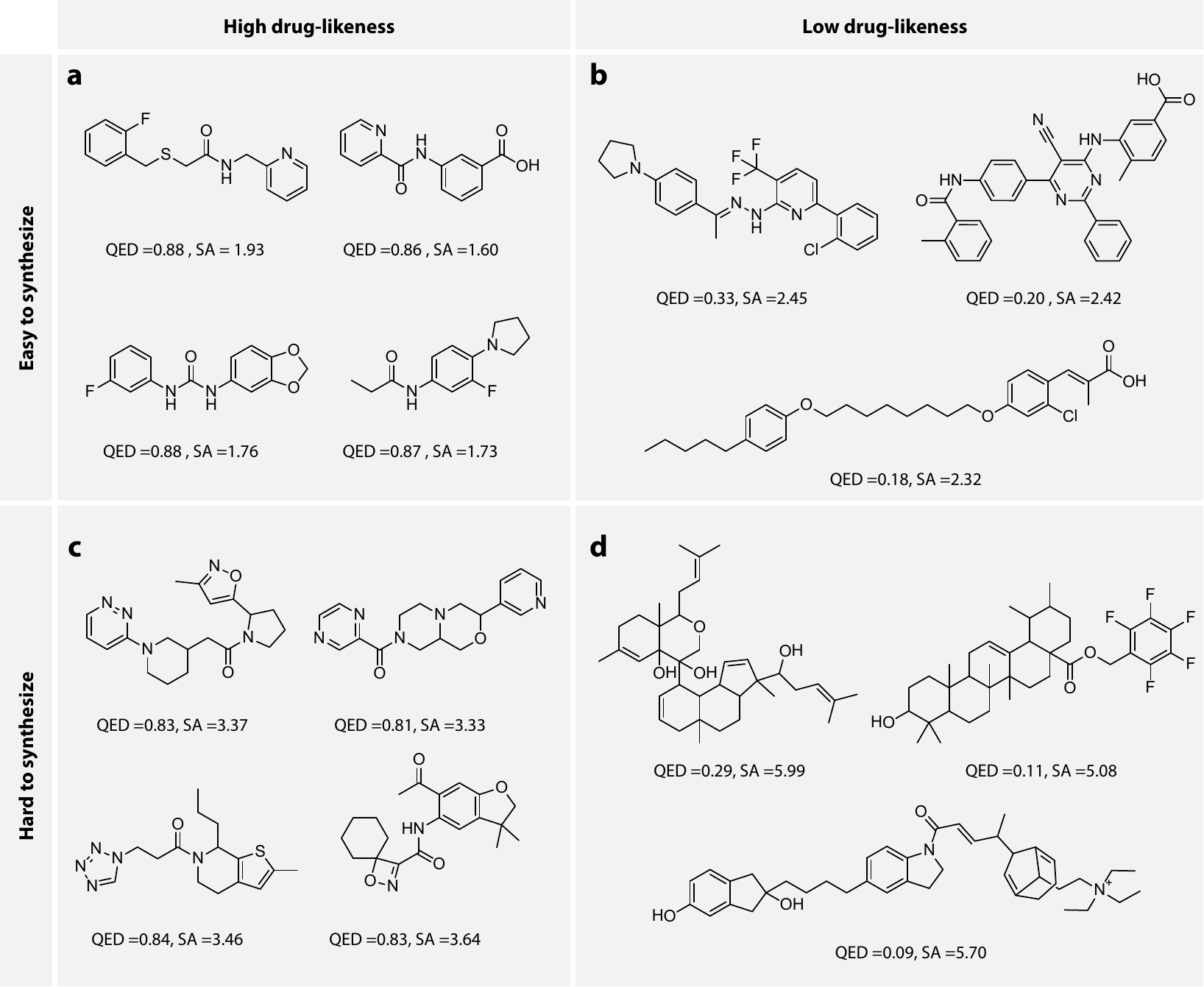}
        \caption{
            \csentence{Samples generated under the four predefined conditions on drug-likeness and synthetic accessibility score}
        }
        \label{fig:prop_samples}
      \end{figure*}

      Quantitative evaluation of graph based and SMILES based models are demonstrated in Table \ref{tab:prop_result}. Again, under all conditions($C_1 \sim C_4$), the graph based model is able to outperform SMILES based model on the rate of valid outputs. The difference is most significant for conditions requiring high SAscore (that is, $C_3$ and $C_4$). This observation suggests that SMILES based model have difficulty in generating complexed structures while maintaining the structural validity. 
      
      The graph based model also provides better performance in terms of $R_C$ and $EOR_C$ as shown in Table \ref{tab:prop_result}. It is noted that both graph and SMILES based models have relatively bad performance on condition $C_3$, which corresponds to molecules with high drug-likeness and low synthetic accessibility. However, this result is easy to understand. Since the definition of drug-likeness contains the requriement for high synthetic accessibility,  finding molecules with high QED score and high SAscore is in itself a difficult task. For other conditions, the $R_C$ results for both models varies from 50\% to 70\%. The values are lower compared with scaffold based task, but nonetheless showing enrichments for all conditions over the distribution from ChEMBL. The diversity of generated samples are also reported. Similar to the observation in \hyperref[sec:scaffold]{``Scaffold-Based Generation''}, SMILES based method is able to produce outputs with slighly higher diversity compared with graph based method.
      
      For a visualized demonstration, the distributions of QED and SA score for the output samples from graph based generator are shown in Figure \ref{fig:prop_distributions}{\bf a-d}. Random samples are also chosen for each class and are visualization in Figure \ref{fig:prop_samples}. The structural features for the output samples are mostly consistent with the predefined conditions, with small and simple molecules for $C_1$ and highly complexed molecules for $C_4$.
      
      Note that conditional model also supports generation based on a given point of QED and SAscore. This possibility is demonstrated for visualization using graph based conditional model. The molecule generation is now conditioned on single points of conditional code ${\bf c}$. Here, we use four different  conditions as specified as follows: ${\bf c}_1=(0.84,1.9)$, ${\bf c}_2=(0.27,2.5)$, ${\bf c}_3=(0.84,3.8)$ and ${\bf c}_4=(0.27,4.8)$.

      \begin{figure*}
        \includegraphics{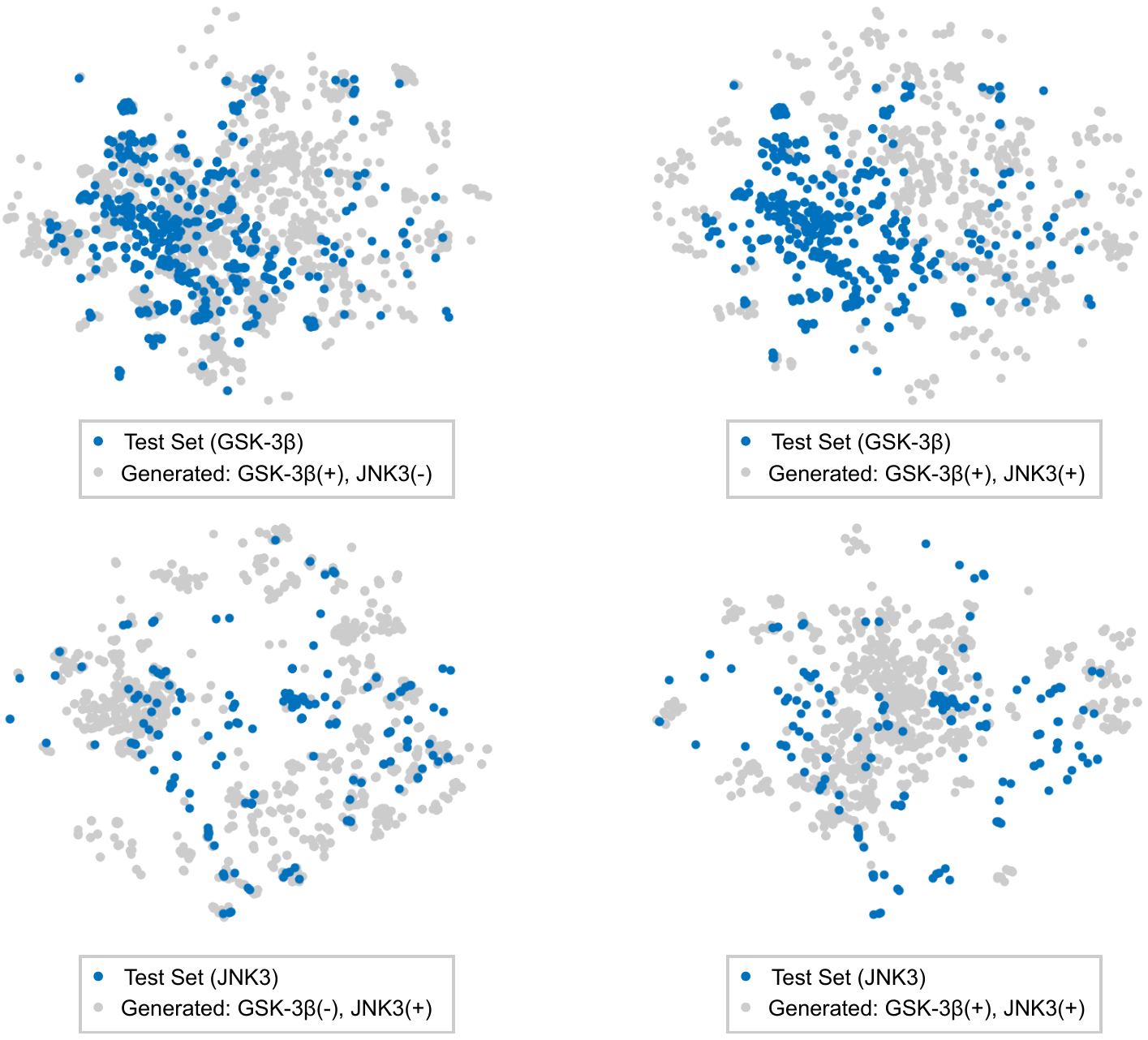}
        \caption{
            \csentence{Visualizing the distribution of generated samples for each target.} The figure shows the t-SNE visualization of: {\bf a}. molecules form test set of GSK-3$\beta$ and samples conditioned on JNK3(-), GSK-3$\beta$(+)  {\bf b}. molecules from test set of GSK-3$\beta$ and samples conditioned on JNK3(+), GSK-3$\beta$(+). {\bf c}. molecules from test set of JNK3 and samples conditioned on JNK3(+), GSK-3$\beta$(-)  {\bf d}. molecules from test set of JNK3 and samples conditioned on JNK3(+), GSK-3$\beta$(+)
        }
        \label{fig:t_sne}
      \end{figure*}

      The distributions of QED and SA for the output molecules by graph based model are shown in Figure \ref{fig:prop_conditions}{\bf e-h}. Results show that although the requirement is specified using a single value of QED and SA score,  the distribution of the two properties for output samples are relatively dispersed. This result is not surprising  since the QED and SA score are relatively abstract descriptions of structural features of molecules, and a small  modification of molecule structure may lead to significant changes in QED and SA scores. Nonetheless, it can be  found that the generated samples are enriched around the corresponding code ${\bf c}$. It is also observed that the  distribution of SA is more concentrated than that of QED. This is probably because that SA is direct measurement of  molecular graph complexity, which may be easier to model for the graph based generator. In contrast, QED is a more  abstract descriptor related to various molecular properties.

    \subsection*{Generating Dual Inhibitors for JNK3 and GSK-3$\beta$}
      In this task, the model is used to generate dual inhibitor for JNK3 and GSK-3$\beta$. A predictive model is first used to label the conditional code for ChEMBL dataset, and the conditional graph generator is trained on the labeled training set. The two predictors yield good results in general, with AUC=0.983 for JNK3 and AUC=0.984 for GSK-3$\beta$. The ROC curves for the two models are show in Figure S4(Additional file 2).

      \begin{table*}
        \caption{Performance of graph based and SMILES based model on inhibitor generation, Results are reported as $Mean \pm StdDev$. The model giving the best performance in each metric is highlighted in boldface}
        \begin{tabular}{ccccccc}
            \toprule
            Condition ($\bf c$) & $R_0$ & Model & \% valid & $R_{\bf c}$ & $EOR_{\bf c}$ & Diversity \\
            \midrule
            \multirow{2}{*}{\begin{tabular}{c}GSK-3$\beta$(+)\\JNK3$\beta$(+)\end{tabular}} & \multirow{2}{*}{$0.0008$} & Graph & $0.939\pm 0.007$ & $0.53\pm 0.01$ & 666 & ${\bf 0.824\pm 0.003}$ \\ 
            &&SMILES&${\bf 0.959\pm 0.003}$ & ${\bf 0.56\pm 0.01}$ & {\bf 697} & $0.820\pm 0.002$ \\ 
            \cmidrule{3-7}
            \multirow{2}{*}{\begin{tabular}{c}GSK-3$\beta$(+)\\JNK3$\beta$(-)\end{tabular}} & \multirow{2}{*}{$0.01$} &Graph &${\bf 0.932\pm 0.007}$ & $0.42\pm 0.01$ & 42 & ${\bf 0.866\pm 0.001}$ \\ 
            &&SMILES&$0.928\pm 0.003$ & ${\bf 0.47\pm 0.01}$ & {\bf 47} & $0.862\pm 0.001$ \\ 
            \cmidrule{3-7}
            \multirow{2}{*}{\begin{tabular}{c}GSK-3$\beta$(-)\\JNK3$\beta$(+)\end{tabular}} & \multirow{2}{*}{$0.0008$} &Graph &${\bf 0.955\pm 0.003}$ & ${\bf 0.61\pm 0.00}$ & {\bf 759} & $0.834\pm 0.001$ \\ 
            &&SMILES&$0.944\pm 0.003$ & $0.56\pm 0.01$ & 698 & ${\bf 0.837\pm 0.001}$ \\ 
            \bottomrule
        \end{tabular}
        \label{tab:kinase_result}
      \end{table*}

      \begin{table*}
        \caption{The $K_{{\bf c}{\bf c}^\prime}$ matrix for kinase inhibitor generation task, the diagnal elements $K_{{\bf c}{\bf c}}=R_{\bf c}$ are omitted since they have been reported in Table \ref{tab:kinase_result}. Results are reported as $Mean \pm StdDev$. The model giving the best performance in each metric is highlighted in boldface}
        \begin{tabular}{ccccc}
            \toprule
            &&\multicolumn{3}{c}{Results(${\bf c}^\prime$)} \\
            \cmidrule{3-5}
            Condition(${\bf c}$) & Model & \begin{tabular}{c}GSK-3$\beta$(+),\\JNK3$\beta$(+)\end{tabular} & \begin{tabular}{c}GSK-3$\beta$(+),\\JNK3$\beta$(-)\end{tabular} & \begin{tabular}{c}GSK-3$\beta$(-),\\JNK3$\beta$(+)\end{tabular} \\
            \midrule
            \multirow{2}{*}{\begin{tabular}{c}GSK-3$\beta$(+),\\JNK3$\beta$(+)\end{tabular}} & Graph & - &$0.178\pm 0.007$ & ${\bf 0.018\pm 0.001} $\\
            &SMILES&- &${\bf 0.167\pm 0.010}$ & $0.063\pm 0.006$ \\
            \multirow{2}{*}{\begin{tabular}{c}GSK-3$\beta$(+),\\JNK3$\beta$(-)\end{tabular}} & Graph &$ {\bf 0.034\pm 0.001} $& - &${\bf 0.003\pm 0.000}$ \\  
            &SMILES&$0.082\pm 0.007$ & - &$0.023\pm 0.002$ \\
            \multirow{2}{*}{\begin{tabular}{c}GSK-3$\beta$(-),\\JNK3$\beta$(+)\end{tabular}} & Graph & ${\bf 0.024\pm 0.004}$ & ${\bf 0.022\pm 0.002}$ & - \\  
            &SMILES&$0.083\pm 0.007$ & $0.057\pm 0.002$ & -\\
            \bottomrule
        \end{tabular}
        \label{tab:kinase_matrix}
      \end{table*}

      Results for both the SMILES based and graph based conditional generator are given in Table \ref{tab:kinase_result}. In terms of output validity, graph based model outperforms SMILES based model in generating GSK-3$\beta$ selective and JNK3 selective compounds, but for the generation of dual inhibitors, SMILES based model outperforms graph based model. In terms of $R_{\bf c}$ and $EOR_{\bf c}$, SMILES based model achieves better performance in the task of generating dual inhibitors and the task of generating selective inhibitors for GSK-3$\beta$, while graph based model achieves better performance in the task of generating JNK3 selective inhibitors.The $K_{{\bf c}{\bf c}'}$ matrices for graph based and SMILES based model are shown in Table \ref{tab:kinase_matrix}. For both graph based and SMILES based model, it is noted that when generating compounds that is active to both JNK3 and GSK-3$\beta$, there is a significant amount of outputs falling into the category of GSK-3$\beta$ positive and JNK3 negative. Nonetheless, in terms of the enrichment over random $EOR_{\bf c}$, the two models are able to achieve high performance for all selectivity combinations. Note that selective inhibitors for GSK-3$\beta$ are relatively enriched in ChEMBL database, according to the result of the predictor. In comparison, the selective inhibitors against JNK3 and the dual inhibitor for both JNK3 and GSK-3$\beta$ are much rarer. However, the model is still able to achieve significant enrichment for the two types of selectivity. The result shows potential application for target combinations that have low data enrichment rate.

      \begin{figure*}[t!]
        \includegraphics{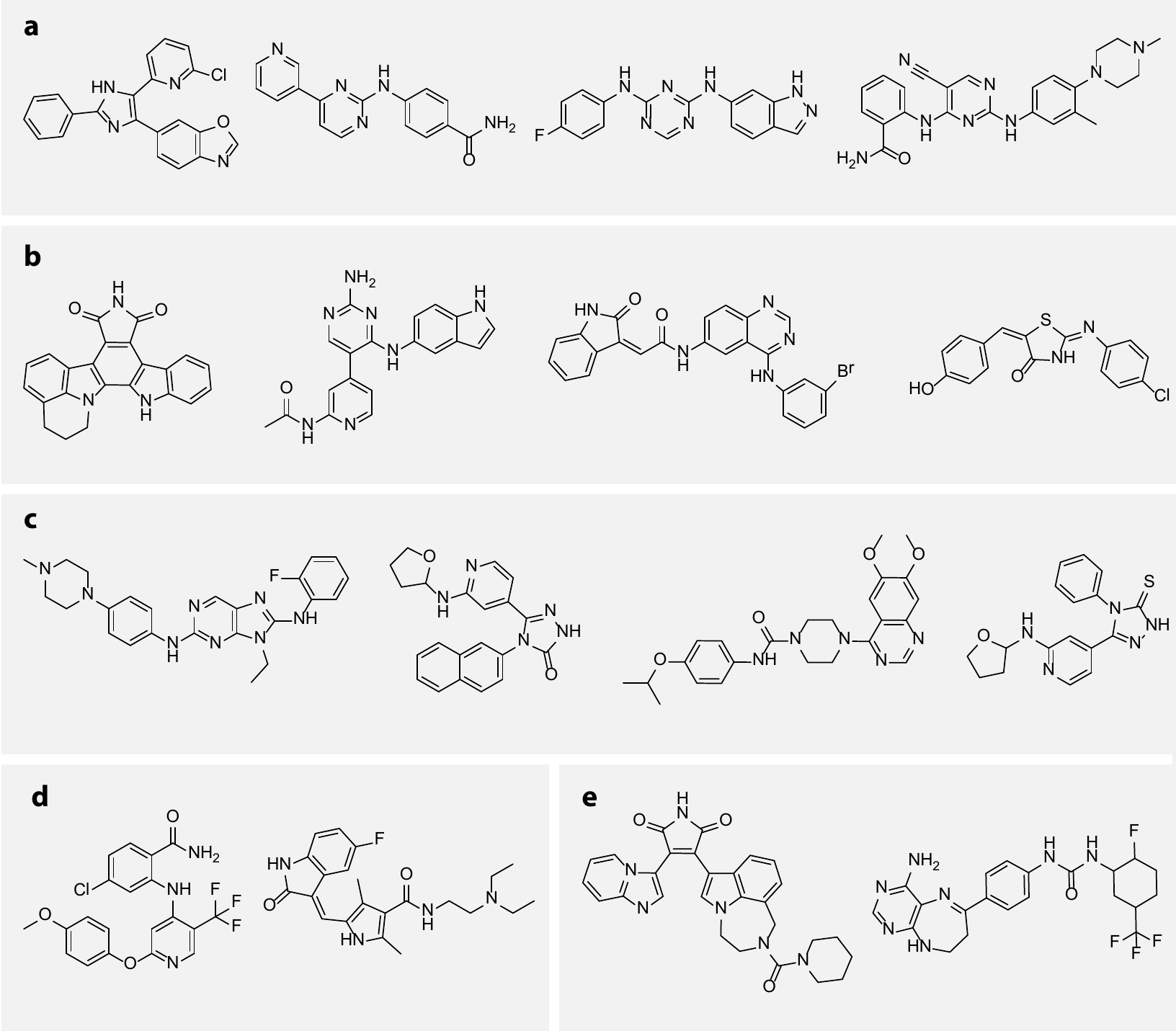}
        \caption{
            \csentence{Samples conditioned on different selectivity conditions.}
            {\bf a-c}. Generated samples under different condition of selectivity, {\bf a} for Dual inhibitors, {\bf b} for GSK-3$\beta$ selective inhibitors, and {\bf c} for JNK3 selective inhibitors.
            {\bf d-e}. Several recovered actives of JNK3(see $\bf d$) and GSK-3$\beta$ (see $\bf e$)
        }
        \label{fig:kinase_samples}
      \end{figure*}
  
      To better demonstrate the structural distribution of the generated samples, visualization based on t-SNE\cite{t_sne}is performed using the ECFP6 fingerprint. The generated samples under different selectivity specifications and molecules in the test set for each target are projected into two-dimensional embeddings and are shown in Figure \ref{fig:t_sne}{\bf a-d}. The result illustrates the structural distribution is well-matched between generated molecules and the test set. It is also shown that the conditional generator tends to produce molecules near the test set samples, which is consistent with observations based on other methods\cite{slm_3}. It is also observed that molecules generated under different selectivity condition occupy distinct region of chemical space.

      For each selectivity condition, several molecules are sampled using the model and are demonstrated in Figure \ref{fig:kinase_samples}{\bf a-c}. By investigating the generated structures in detail, it can be observed that the model tends to generate samples containing well-established scaffold for the corresponding target. For JNK3, structures such as diaminopurines\cite{jnk_1}  and triazolones\cite{jnk_2}, which has frequently been used in the design of JNK inhibitors, show high occurrence in  the generated samples. This observation is the same for GSK-3$\beta$, with example like 2,3-bis-arylmaleimides, a  class of widely studied inhibitor of GSK-3\cite{review_jnk_2}. On the other hand, aminopyrimidines are frequently  shown in the outputs of all selectivity conditions, but they are more enriched in generated dual inhibitors. Those  observations show good interpretability of the outputs, and indicate that the structural features of generated  samples are in line with the existing knowledge about the two targets.

      Finally, we report the percentage of reproduced samples from the test set for each target. From the result, 10.3\% of molecules are reproduced for JNK3 and, 6.0\% of molecules are reproduced for GSK-3$\beta$. Note that molecules in the test sets for each targets have been excluded from the ChEMBL training set in this task, which means that the method is capable of generating molecules that have been confirmed to be positive, without seeing them in the training set of predictive model and conditional generative model. 

      Several recovered actives are shown in Figure \ref{fig:kinase_samples}{\bf d}-{\bf e}. Those molecules show relatively high diversity in structure, indicating that the model does not collapse to a subgroup of active compounds. A quantitative evaluation is performed using the internal diversity, and the result shows that the recovered GSK-3$\beta$ inhibitors have a internal diversity of 0.819, while the recovered JNK3 inhibitors have a internal diversity of 0.761. Those values are relatively close to the diversity of test set molecules, which are 0.867 for GSK-3$\beta$ and 0.852 for JNK3.

  \section*{Conclusion}
    In this work, a new framework for {\it de novo} molecular design is proposed based on graph generative model and is  applied to solve different drug design problems. The graph generator is designed to be more fitted to the tasks  of molecule generation by using a simple decoding scheme and a graph convolutional architecture that is less  computationally expensive. Furthermore, a more flexible way of introducing decoding invariance is also suggested.  The method is trained using molecules in ChEMBL dataset and has been demonstrated to have better performance compared with SMILES based methods, especially in terms of the rate of valid outputs.

    To generate molecules with specific requirements, we propose to use conditional generative model, which provides  higher flexibility and is much easier to train compared with previous fine-tuning based methods. The model  is applied to solve problems that is highly related to drug design, such as generating molecules based on a given  scaffold, generating molecules with good drug-likeness and synthetic accessibility and the generation of  molecules with specific profile against multiple targets. The high enrichment rates presented in the results  show that the conditional generative model provides a promising solution for many real-life drug design tasks.

    This work can be extended in various aspects. First of all, the models used in this work completely ignores  the stereochemistry information for molecules. In fact, stereochemistry is extremely important in the process  of drug development, and introducing this information helps to improve the applicability of existing models.  Secondly, for the target based generation, it will be much more helpful to jointly train the generator and the  decoder, utilizing strategies such as semi-supervised learning\cite{semi_1, semi_2}. Finally, besides the three tasks  experimented in this work, conditional graph generator can be used in many other scenarios. To summarize,  the graph generative architecture proposed in this work gives promising result in various drug design tasks,  and it is worthwhile to explore other potential applications using this method.
 
  \begin{backmatter}

      
      \section*{Additional Files}
        \subsection*{Additional file 1 --- Supplementary Text}
          Containing additional information about the model architecture and implementation details of experiments.
      
        \subsection*{Additional file 2 --- Supplementary Figures}
          Contianing supplementary figures.
      
      \section*{Availability of data and materials}
        The source code and data supporting the conclusions of this article is available at 
        \url{https://github.com/kevinid/molecule_generator}.
      
      \section*{List of abbreviations}
      \begin{itemize}
      \item{SMILES - Simplified molecular-input line-entry system}
      \item{RNN - Recurrent neural network}
      \item{LM - Language model}
      \item{RF - Random forest}
      \item{RL - Reinforcement learning}
      \item{VAE - Variational autoencoder}
      \item{GRU - Gated recurrent unit}
      \item{DRD2 - Dopamine receptor D2}
      \item{JNK3 - c-Jun N-terminal kinase 3}
      \item{GSK3$\beta$ - glycogen synthase kinase-3 beta}
      \item{QED - Quantitative estimate of drug-likeness}
      \item{SA - Synthetic accessibility}
      \item{ECFP - Extended connectivity fingerprint}
      \item{t-SNE - t-Distributed stochastic neighbor embedding}
      \end{itemize}
      
      \section*{Competing interests}
        The authors declare that they have no competing interests.
      
      \section*{Author's contributions}
      Yibo Li formulated the concept and contributed to the implementation. 
      Yibo Li wrote the manuscript, Liangren Zhang and Zhenming Liu reviewed and edited the manuscript. 
      All authors read and approved the final manuscript.
      
      \section*{Acknowledgements}
      We would like to thank Xiaodong Dou for his help on the discussion of generated inhibitors of JNK3 and GSK3$\beta$. Thanks to Bo Yang who helped with the profiling of Supplementary Text 8.
      
      \section*{Funding}
      This  research was supported by the National Natural Science Foundation of China (Grant 81573273, 81673279, 21572010 and 21772005) as well as National Major Scientific and Technological Special Project for ``Significant New Drugs Development'' (Grant 2018ZX09735001-003)
      
      \bibliographystyle{bmc-mathphys}
      \bibliography{bmc_article}

  \end{backmatter}

\end{document}